\documentclass[journal]{IEEEtran} 

\usepackage[labelsep=colon, singlelinecheck=false]{caption}
\usepackage{amsmath,amsfonts}
\usepackage{algorithmic}
\usepackage{algorithm}
\usepackage{array}
\usepackage[caption=false,font=normalsize,labelfont=sf,textfont=sf]{subfig}
\usepackage{textcomp}
\usepackage{stfloats}
\usepackage{url}
\usepackage{verbatim}
\usepackage{graphicx}
\usepackage{cite}
\usepackage{comment}

\usepackage{amssymb}
\usepackage{color}
\usepackage{xcolor}
\usepackage{hyperref}

\usepackage{enumerate}

\usepackage{amsthm}

\AfterEndEnvironment{theorem}
{\noindent\ignorespaces}

\AfterEndEnvironment{Definition}{\noindent\ignorespaces}

\AfterEndEnvironment{lemma}{\noindent\ignorespaces}
\newtheorem{proposition}{Proposition}
\AfterEndEnvironment{proposition}{\noindent\ignorespaces}

\def\x{{\mathbf x}}

\def\y{{\mathbf y}}
\def\c{{\mathbf c}}
\def\a{{\mathbf a}}

\def\r{{\mathbf r}}
\def\f{{\mathbf f}}
\def\g{{\mathbf g}}

\def\s{{\mathbf s}}

\def\ubf{{\mathbf u}}
\def\vbf{{\mathbf v}}

\def\Acal{{\mathcal A}}
\def\Dcal{{\mathcal D}}
\def\X{{\mathcal X}}
\def\H{{\mathbb H^d}}
\def\K{{\mathbb R^p}}
\def\Hcal{{\mathcal H}}
\def\Kcal{{\mathcal K}}

\def\Tbf{{\mathbf T}}
\def\Ncal{{\mathcal N}}
\def\Qcal{{\mathcal Q}}
\def\Ccal{{\mathcal C}}

\def\S{{\mathbf S}}
\def\St{{\tilde{\mathbf S}}}
\def\Scal{{\mathcal S}}
\def\Rcal{{\mathcal R}}
\def\C{{\mathbb C}}
\def\R{{\mathbb R}}
\def\U{{\mathbf U}}
\def\V{{\mathbf V}}
\def\P{{\mathbf P}}

\def\F{{\mathbf F}}
\def\Ft{{\tilde{\mathbf F}}}
\def\G{{\mathbf G}}
\def\A{{\mathbf A}}

\def\bigPhi{{\mathbf \Phi}}
\def\bigPsi{{\mathbf \Psi}}

\def\dtt{{\text{sim}}}

\def\I{{\mathbf I}}

\newcommand{\spann}[1]{\text{span}(#1)}

\newcommand{\inn}[2]{\langle {#1}, {#2}\rangle}

\newcommand{\constraint}[3]{%
  &(#1) \; #2 \quad #3 \nonumber
}

\newcommand{\constraintend}[3]{%
  &(#1) \; #2 \quad #3
}

\newcommand{\conprd}{%
\constraint{c}{f_k^{\textrm{MIN}} < f_k^c(t) \leq f_k^{\textrm{MAX}}}{\forall\, k,\, t;}\\
\constraint{d}{{f^r}^{\textrm{MIN}} < {f^r}(t) \leq {f^r}^{\textrm{MAX}}}{\forall\, t;}\\
\constraint{e}{R_k^{\textrm{MIN}} \leq R_k(t) \leq R_k^{\textrm{MAX}}(t)}{\forall\, k,\, t;}\\
\constraint{f}{B_k(t)\geq 0}{\forall\, t,k;}\\
\constraint{g}{\sum_{k=1}^{K} {B_k}(t) = {B}}{\forall\, t;}\\
\constraint{h}{q_k(t)\in \mathcal{Q}}{\forall\, k,\, t;}\\
\constraintend{i}{N_k(t) \in \mathcal{N}}{\forall\, k,\, t;}
}

\begin{document}

\title{Frame-Based Zero-Shot Semantic Channel Equalization for AI-Native Communications}

\author{Simone Fiorellino, Claudio Battiloro, Emilio Calvanese Strinati, and Paolo Di Lorenzo 
\thanks{Simone Fiorellino is with the Department of Computer, Control, and Management Engineering (DIAG), Sapienza University of Rome, via Ariosto 25, 00185 Rome, Italy, and also with CNIT, Parma, Italy (e-mail: simone.fiorellino@uniroma1.it). Claudio Battiloro is with the Biostatistics Department of Harvard University, 655 Huntington Ave, Boston, MA 02115, USA (e-mail: cbattiloro@hsph.harvard.edu). Emilio Calvanese Strinati is with CEA Leti, University Grenoble Alpes, 38000 Grenoble, France (e-mail: emilio.calvanese-strinati@cea.fr). Paolo Di Lorenzo is with the Department of Information Engineering, Electronics, and Telecommunications (DIET), Sapienza University of Rome, 00184 Rome, Italy, and also with CNIT, Parma, Italy (e-mail: paolo.dilorenzo@uniroma1.it). 
This work was supported by the European Union under the Italian National Recovery and Resilience Plan (NRRP) of NextGenerationEU, partnership on “Telecommunications of the Future” (PE00000001 - program “RESTART”), by the Next Generation EU, Mission I.3.3 PNRR Scholarships for Innovative PhD Programmes Addressing Business Innovation Needs, Component 2, CUP 352: B53C22001970004, and by the SNS JU project 6G-GOALS under the EU’s Horizon program Grant Agreement No 101139232. A preliminary conference version of this work appeared in \cite{fiorellino2024dynamic}.}}



\maketitle

\begin{abstract}
In future AI-native wireless networks, the presence of mismatches between the latent spaces of independently designed and trained deep neural network (DNN) encoders may impede mutual understanding due to the emergence of semantic channel noise. This undermines the receiver’s ability to interpret transmitted representations, thereby reducing overall system performance. To address this issue, we propose the Parseval Frame Equalizer (PFE), a zero-shot, frame-based semantic channel equalizer that aligns latent spaces of heterogeneous encoders without requiring system retraining. PFE enables dynamic signal compression and expansion, mitigating semantic noise while preserving performance on downstream tasks. Building on this capability, we introduce a dynamic optimization strategy that coordinates communication, computation, and learning resources to balance energy consumption, end-to-end (E2E) latency, and task performance in multi-agent semantic communication scenarios. Extensive simulations confirm the effectiveness of our approach in maintaining semantic consistency and meeting long-term constraints on latency and accuracy under diverse and time-varying network conditions.
\end{abstract}


\section{Introduction}

\IEEEPARstart{M}{odern} wireless communication systems are being pushed to their limits as the number of connected devices and the volume of data traffic rise. Traditional bit-centr schemes, often reliant on bandwidth scaling for higher throughput, are becoming less practical under stringent spectrum availability, rising energy consumption, and constrained computational resources at the network edge \cite{alwis2021devices,shi2021new,li2018Beyond,saad2019vision}. 
Data-intensive applications such as autonomous driving, industrial automation, and smart healthcare highlight the need for a more efficient and context-driven framework that effectively reduces network overhead and latency while maintaining reliability and adaptability.
Consequently, researchers in academia and industry alike are converging on the concept of semantic communications (SCs) \cite{shannon1948mathematical,strinati20216g,strinati2024goal,lu2022rethinking,xie2021deep,gunduz2022beyond,getu2023semantic}. SCs focus on conveying the meaning or intent behind the transmitted bits and go beyond the exact reconstruction of the raw data. This shift is largely driven by artificial intelligence (AI), particularly DNNs, where E2E communication solutions learn to extract, transmit, and decode high-level semantic features, enabling more efficient use of bandwidth and energy. One important direction within the broader SC paradigm is the \emph{goal-oriented} (GO) \emph{communication}, wherein the primary objective is the fulfillment of a task (e.g., learning, control, actuation, etc.) characterized by a set of system requirements \cite{di2023goal}, thus enabling a GO data compression \cite{zhang2023goal,stavrou2023role}. This paradigm is particularly relevant in scenarios where agents (whether human or artificial) must act upon received information, relaxing the necessity of data reconstruction.

Building on this idea, many methodologies have emerged in the literature \cite{ gunduz2022beyond, xie2021deep,stavrou2023role,di2023goal, binucci2023multi, merluzzi20246g, shao2021learning}. 
For example, \cite{kountouris2021semantics} proposes semantics-driven sampling and communication policies for a real-time source reconstruction scenario involving remote actuation, achieving significant reductions in reconstruction error, actuation costs, and redundant data transmission. Interestingly, \cite{shao2021learning} employs a variational information bottleneck to learn compact, bitrate‑adaptive feature encodings optimized for low‑latency edge inference under bandwidth constraints. Other GO frameworks for adaptive quantization, compression, and resource allocation have been proposed in \cite{merluzzi2021wireless, shlezinger2021deep, binucci2022dynamic,binucci2022adaptive,di2023goal} to explore optimal trade‑offs among key performance indicators (KPIs) in wireless edge intelligence. 
Deep generative models have also been integrated into SCs to enable GO content generation across modalities: for text‑to‑image \cite{nam2024language}, for image \cite{han2023generative,barbarossa2023semantic, grassucci2023generative}, and for audio \cite{jang2024personalized, grassucci2024diffusion}.
Another approach is deep joint source-channel coding (JSCC) \cite{jankowski2020wireless}, in which the feature vectors learned by DNN encoders are directly mapped into channel input symbols.

Introducing semantics into communication systems brings new challenges, notably in the form of \emph{semantic noise}, which arises from inconsistencies in logic, interpretation, or background knowledge among AI agents \cite{getu2023semantic,luo2022semantic}. For AI-native communication to be effective, both the transmitter and the receiver must share a common, or compatible, latent representation space. However, due to differences in network architectures or training processes, devices often produce divergent latent representations for the same underlying information. This misalignment introduces semantic noise, potentially degrading mutual understanding between agents. Such inconsistencies are common in multi-vendor environments, where sharing models, training datasets, or proprietary assets is either infeasible or undesirable. In these contexts, joint end-to-end training or model exchange is often ruled out due to privacy concerns and intellectual property restrictions. Therefore, the development of robust semantic alignment mechanisms is essential to ensure consistency and interoperability across heterogeneous systems.

\textbf{Related works.} In this paper, we focus on latent space mismatch as a form of semantic noise. More broadly, the problem of estimating correspondences between two latent spaces is well-known and can be referred to, for instance, as manifold alignment \cite{wang2009manifold}, feature set matching \cite{grauman2005pyramid}, or feature correspondence finding \cite{torresani2008feature}.
Specifically, we treat the problem as a case of \emph{Semantic Channel Equalization} (SCEq) \cite{sana2023semantic}, to \emph{equalize} the semantic discrepancy between a transmitter and a receiver in an AI-native communication system. For instance, in \cite{sana2023semantic}, the authors propose an algorithm that relies on a codebook of low-complexity linear transformations that map between the \textit{atoms} (i.e., regions of shared semantic meaning) in the transmitter and receiver latent spaces, governed by a suitable selection policy. Although this method has demonstrated its effectiveness, it requires explicitly identifying and defining these atoms across different models, which is not straightforward. In \cite{huttebraucker2024soft}, this challenge was addressed by defining the atoms in a self-supervised manner, which improved overall performance of the equalization algorithm. Nevertheless, the process still relies on labeled data for atom identification, and the transformation selection policy requires precise atom estimation. 
Recently, in \cite{moschella2022relative}, it was introduced the \emph{Relative Representations} (RRs) framework to establish a common semantic representation between distinct (independently trained) DNN-based encoding functions in a zero-shot fashion, i.e., eliminating the need for additional optimization steps and even the exchange of latent representations, as typically required in supervised settings. Specifically, rather than relying on absolute latent representations (i.e., the direct output of a DNN-based encoder), RRs map the data embeddings into a \emph{relative space}, where each component corresponds to a similarity score between the embeddings of the desired input and the embeddings of a set of predefined \emph{anchors}. The effectiveness of RRs in unifying latent spaces has been demonstrated in text and image applications \cite{moschella2022relative}, reinforcement learning \cite{ricciardi2023zero}, and across various similarity metrics \cite{cannistraci2023bricks}, with generalization capabilities highlighted in \cite{crisostomi2023charts,norelli2023asif}.  Furthermore, \cite{fiorellino2024dynamic} presents a dynamic optimization strategy that employs relative representations to adapt communication, computation, and learning resources, such as anchor sets and encoders, enabling energy-efficient, low-latency, and effective semantic communications. Although most existing studies assume that a relative decoder can be trained in the new representation space, the training process itself may be infeasible in some scenarios (e.g., resource-constrained networks). To overcome this limitation, \cite{maiorca2024latent} leverages a cosine-similarity-based relative representation, which admits an invertible mapping for alignment across models without necessitating the training of an additional decoder. Instead, a sample-efficient optimization algorithm is used in \cite{huttebraucker2024relative} to achieve absolute representation reconstruction, thereby generalizing SCEq to arbitrary similarity functions.

\textbf{Contributions. }
In this work, we investigate zero-shot semantic channel equalization in multi-agent systems by focusing on how a receiver (RX) can reconstruct its original latent representation (i.e., the absolute space) from a set of coefficients transmitted by a non-aligned (unknown) transmitter (TX). Specifically, building on \cite{maiorca2024latent, huttebraucker2024relative}, we leverage the zero-shot stitching capability of RRs to introduce a semantic channel equalization mechanism integrated with a Parseval frame-based system. The properties of Parseval frames permit linear and well-conditioned pre- and post-semantic equalizers, facilitate transmission of \textit{frame coefficients}, and ensure low computational overhead and feasibility in communication systems.
Varying the number of transmitted coefficients lets the same mechanism control \textit{compression} (reducing semantic payload through orthogonal projection) and \textit{redundancy} (enhancing robustness via frames). This design achieves an optimal rate–reliability trade-off in the semantic channel.
To further capitalize on this trade-off, we extend the dynamic optimization framework proposed in \cite{fiorellino2024dynamic} to a multi-agent setting using Lyapunov stochastic optimization \cite{2010Neely}. The system automatically balances semantic channel equalization performance with frame coefficient compression and quantization, thereby dynamically allocating communication, computation, and learning resources. The resulting architecture fosters an energy-efficient, low-latency semantic communication paradigm with robust semantic consistency across heterogeneous encoders. Importantly, unlike \cite{fiorellino2024dynamic}, which requires a set of relative decoders at the receiver to interpret transmitted relative representations, the frame-based equalization reconstructs the absolute latent representation directly, obviating the need for additional relative decoders at the RX. Numerical simulations demonstrate that our approach consistently preserves semantic consistency while meeting long-term latency and accuracy requirements, even under dynamic and heterogeneous network conditions.

\textbf{Notation.} Scalar, column vector, and matrix variables are respectively indicated by plain letters $a$ ($A$), bold lowercase letters $\a$, and bold uppercase letters $\A$. The $n$-th component of a vector is indicated by $[\a]_n$. We will refer to sets (and latent spaces) with calligraphic uppercase letters $\mathcal{A}$. Additional notation is introduced as needed throughout the paper.

\section{Background}
\label{sec:background}

In this section, we provide an overview of the two main building blocks for the design of the proposed semantic channel equalizer: (i) relative representations, a zero-shot latent-space stitching technique; and (ii) frame theory, which provides well-conditioned linear operators.

\subsection{Relative Representations}
\label{sec:background:relrep}

Consider two distinct encoding functions, $E_\Hcal$ and $E_\Kcal$, both trained on the same dataset $\Dcal$. Each function maps an input signal $\s \in \Dcal$ to a corresponding lower-dimensional space, with $E_\Hcal: \Dcal \to \mathbb{R}^d$ and $E_\Kcal: \Dcal \to \mathbb{R}^p$ denoting the respective encoders. For clarity, we denote the set of all latent vectors embedded by a generic function $E_{(\cdot)}$ as $\X_{(\cdot)}$. In particular, we identify $\mathbb{R}^d$ with $\X_{\Hcal}$ and $\K$ with $\X_{\Kcal}$, thereby clearly distinguishing the latent spaces generated by each encoding function.
The objective of the RRs is to project the $\X_\Hcal$ and $\X_\Kcal$ in a unified relative space. To build this space, we select a subset $\Acal$ of $\Dcal$ of cardinality $N$ denoted as \textit{anchors set}, i.e., $\Acal=\{s_1,...,\s_N\}\subset\Dcal$. 
For the latent space $\X_\Hcal$, the RR of a point $\x\in\X_\Hcal$ is given by \cite{moschella2024latent}:
\begin{align}\label{eq:def_RR}
    \r_{\s}= [\dtt(\x,\a_1),\dtt(\x,\a_2),\ldots,\dtt(\x,\a_{N})]^T\in\R^{N}
\end{align}
where $\dtt$ denotes a similarity function and $\a_n$ are the anchor latent representations, i.e., $\a_n=E_{\Hcal}(\s_n)$. Intuitively, the resulting vector $\r_{\s}$ represents a new set of coordinates for $\x$, defined relative to the fixed anchor set $\Acal$. We define the collection of all $\r_\s$ ($\forall\;\s\in\Dcal$) as $\Rcal_\Hcal(\Dcal,\Acal,\dtt)$. 

\textbf{Remark 1.}
If the similarity function in \eqref{eq:def_RR} is the cosine similarity ($S_C$), the resulting RRs depend solely on pairwise angles between embeddings. Thus, the RR is invariant under any angle–norm preserving transformation. As a consequence, two encoders $E_\Hcal$ and $E_\Kcal$ that differ by such a transformation will yield approximately equivalent RRs:
\begin{align}\label{eq:relrep_similarity}
    \Rcal_\Hcal(\Dcal,\Acal,S_C) \approx \Rcal_\Kcal(\Dcal,\Acal,S_C).
\end{align}
The invariance property in \eqref{eq:relrep_similarity} justifies using RRs as projections into a unified semantic space $\Rcal$, defined by a shared anchor set $\Acal$ and similarity function $S_C$. Hence, a single decoder $D:\mathcal{R} \to \R^T$ can be trained to perform a $T$-class classification task directly from the relative space, thus allowing seamless module composition and zero-shot generalization across different encoders.

\textbf{Remark 2.}
\cite{maiorca2024latent} shows that, using $S_C$ as the similarity function, the absolute representation $\x$ can be approximately recovered from its relative form $\r_\s$, avoiding task-specific relative decoders, via an inverse transformation. Specifically, let $\A \in \R^{N \times d}$ be the anchor matrix with rows as latent anchor vectors. Using the Moore–Penrose pseudoinverse $\A^\dagger = (\A^T\A)^{-1}\A^T$, an estimate of $\x$ is given by $\hat{\x} = \A^\dagger \r_\s$.
Alternatively, \cite{huttebraucker2024relative} introduces a sample-efficient, optimization-based inversion that generalizes this recovery to arbitrary similarity functions $\dtt$, extending semantic equalization beyond cosine similarity.

\subsection{Frame Theory}
\label{sec:background:frame}

A set of vectors $\{\f_n\}_{n=1}^{N}$ in an $d$-dimensional Hilbert space $\H$ over the complex field $\C$ is called a \textit{frame} if there exist constants $A > 0$ and $B < \infty$ such that \cite{waldron2018introduction}:
\begin{align}
    A \|\x\|^2 \leq \sum_{n=1}^N |\langle \x, \f_n \rangle|^2 \leq B \|\x\|^2, \quad  \hbox{for all\;} \x \in \H.
\end{align}
The constants $A$ and $B$ are known as the \textit{frame bounds}. If $A = B$, the frame is called a \textit{tight frame}. If $A = B = 1$, the frame is called a \textit{Parseval frame} (PF). The \textit{analysis operator} of the frame is the operator $\F:\H\to \ell_2^N$ given by $\F\x=\{\inn{\x}{\f_n}\}_{n=1}^{N}$, where $\ell_2^{N}\;:=\;\bigl\{\,\mathbf{c}\in\C^{N}\,:\,\sum_{n=1}^{N}\bigl|[\c]_{n}\bigr|^{2}<\infty\bigr\}$ and $\langle\,\cdot\,,\,\cdot\,\rangle$ the canonical inner product . The adjoint of $\F$ is called the \textit{synthesis operator} $\F^H:\ell_2^N\to\H$.
Instead, the \textit{frame operator} $\S: \H \to \H$ is defined as:
\begin{align}\label{eq:frame-operator}
    \S\x = \sum_{n=1}^N \langle \x, \f_n \rangle \f_n.
\end{align}
In matrix form, with $\F$ the matrix whose rows are frame vectors, $\S$ can be expressed as $\S = \F^H\F\in \C^{d\times d}$, where $\S$ is positive, self-adjoint, and invertible \cite{waldron2018introduction}. In finite dimensions, frames coincide with spanning sets:
\begin{proposition}$\!$\cite[Prop. 3.18]{han2007frames}
    Suppose that $\H$ is a finite-dimensional Hilbert space and $\{\f_n\}^{N}_{n=1}$ is a finite collection of vectors from $\H$. Then the following statements are equivalent:
    \begin{enumerate}[(i)]
        \item $\{\f_n\}^{N}_{n=1}$ is a frame for $\H$,
        \item $\spann{\{\f_n\}^{N}_{n=1}} = \H$.
    \end{enumerate}
\end{proposition}
Here
$    
\spann{\{\f_n\}_{n=1}^{N}}
:=\Bigl\{\,\sum_{n=1}^{N}\alpha_{n}\f_{n}\,\Bigm|\,
 \alpha_{n}\in\C\Bigr\}
$. Hence, for finite dimensions, having a frame simply means we have enough vectors to span the whole original space, possibly with overlaps. This redundancy is the key advantage of frames in real-world applications, such as communications, because it adds robustness against noise.

Given \eqref{eq:frame-operator}, any frame reconstructs a signal via
\begin{align}\label{eq:recon_formula}
    \x = \sum_{n=1}^N \langle \x, \f_n \rangle \S^{-1} \f_n.
\end{align}
The numerical stability of computing $\S^{-1}$ depends on the condition number of $\S$, defined as $\text{cond}(\S)=\frac{B}{A}={\text{cond}(\F)}^2\geq1$. Large values, therefore, translate into severe error magnification, especially in high-dimensional spaces. Tight frames offer an elegant solution, having $A=B$, the condition number collapses to one, and the inversion is numerically stable.

A particularly important class of tight frames is the \emph{Parseval frames} (PFs), whose  reconstruction simplifies to a pure inner–product expansion:
\begin{proposition}\cite[Prop. 3.11]{han2007frames}
Given a frame $\{\f_n\}^{N}_{n=1}$ is a Parseval frame for a Hilbert space $\H$ if and only if the following formula holds:
\begin{align}\label{eq:parseval_recon_formula}
    \x = \sum_{n=1}^N \langle \x, \f_n \rangle \f_n, \quad \hbox{for all $\x\in\H$.}
\end{align}
\end{proposition}
Eq. \eqref{eq:parseval_recon_formula} is called the Parseval Reconstruction Formula (PRF). 
Importantly, any frame can be turned into a Parseval frame by “whitening” it with the inverse square root of its frame operator:
\begin{proposition}\label{th:parseval}\cite[Prop. 3.31]{han2007frames}
    Let $\{\f_n\}_{n=1}^N$ be a frame for $\H$ with frame operator $\S$. Then $\{\S^{-1/2} \f_n\}_{n=1}^N$ is a Parseval frame for $\H$.
\end{proposition}
In matrix form, the corresponding Parseval frame matrix is $\Ft \;=\; \F {\S}^{-\tfrac{1}{2}}$, and
yields a new operator $\tilde{\S} = \Ft^H\Ft$ satisfying $\tilde{\S} = \I_d$. 
Consequentially, $\Ft$ is perfectly conditioned and the signal reconstruction \eqref{eq:recon_formula} simplifies in \eqref{eq:parseval_recon_formula}.

\section{Semantic Channel equalization via Parseval Frames}
\begin{figure*}[t!]
    \centering
    \includegraphics[width=\textwidth, trim=0bp 0bp 12bp 0bp, clip]{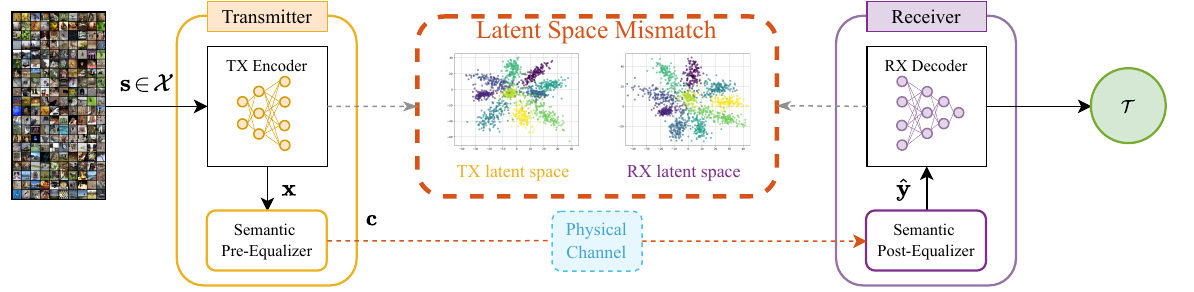}
    \vspace{-.2cm}
    \caption{Diagram of semantic communication enabled by latent space alignment.} 
     \vspace{-.2cm}
    \label{fig:semcom_system_model}
\end{figure*}
We consider a wireless communication system in which both the transmitter and the receiver employ independently pre-trained DNNs to process semantic information. An input signal $\s\in \Dcal$ (e.g., an image or sensor reading) is encoded at the TX into latent vectors $\x\in\R^d$, and the set of all such vectors defines the \textit{TX latent space.} The RX, however, is designed with a different training pipeline and, therefore, its decoder operates on latent vectors $\y\in\R^p$ drawn from its own \textit{RX latent space}, whose dimension $p$ may differ from $d$. 

Because the two latent spaces originate from distinct encoding schemes, a vector produced by the TX does not, in general, belong to the distribution expected by the RX decoder. This \textit{latent-space mismatch} appears at the semantic level and behaves like an additional noise source, commonly referred to as \textit{semantic noise} \cite{sana2023semantic}. The effect is to distort the meaning of the transmitted information even in the absence of physical channel impairments.
To ensure reliable interpretation at the RX, the system incorporates a SCEq stage (semantic pre/post-equalizer) that conceptually realigns TX latents with the structure expected by the RX decoder. By suppressing semantic noise through this adaptation, end-to-end intelligibility of the transmitted content is restored. A pictorial overview is provided in Fig. \ref{fig:semcom_system_model}. In the sequel, we will introduce our SCEq scheme based on Parseval frames.

\subsection{Parseval Frame Equalizer}
\label{sub:PFE}

Our approach to SCEq is based on the idea of RR introduced in Section \ref{sec:background:relrep}, empowered with frame-theoretic arguments. As for RR, we start with the notion of a \emph{anchor set}, $\mathcal{A}_N = \{ s_1, \dots, s_N \} \subset \Dcal$, which is a collection of fixed data samples shared across agents. This common reference set enables each agent to derive its own (private), well-conditioned frame-theoretic analysis operator, denoted by $\{\mathbf{f}_n\}_{\Acal_N} \in \mathcal{X}_{(\cdot)}$, corresponding to the semantic representations of the anchor set $\Acal_N$. 
In other words, our approach treats the anchor matrix $\A$ introduced in Sec. \ref{sec:background:relrep} as an \emph{analysis operator} $\F$ of a frame. The main advantage of our approach is the numerical stability of the reconstruction process. Indeed, if $\F$ forms a \emph{tight} frame, better yet a \emph{Parseval} frame, then $\S=\I$ and the pseudoinverse reduces to $\F^{\!H}$, perfectly conditioned and trivially computable. Conversely, any full-rank but non-tight $\F$ can be whitened into a Parseval frame $\widetilde{\F}= \F\,( \F^{\!H}\F)^{-1/2}$, guaranteeing $\widetilde{\F}^{\,H}\widetilde{\F}= \I$ and condition number equal to 1.

Let $\mathcal{X}_\mathcal{H}$ and $\mathcal{X}_\mathcal{K}$ denote the latent spaces of the TX and RX, respectively, each equipped with its own private analysis operator $\{\mathbf{f}_n\}_{\Acal_N} \subset \mathcal{X}_\mathcal{H}$ at the TX and $\{\mathbf{g}_n\}_{\Acal_N} \subset \mathcal{X}_\mathcal{K}$ at the RX. Let $\mathbf{x} \in \mathcal{X}_\mathcal{H}$ and $\mathbf{y} \in \mathcal{X}_\mathcal{K}$ represent the latent representations of the same semantic input $\mathbf{s} \in \Dcal$, as encoded by the TX and decoded by the RX. 
W.l.o.g., let us assume that the latent spaces generated by the encoding functions are of equal dimension, $d=p$. When $d\neq p$ we choose $\Acal_N$ such that it has at least $N=\max(d,p)$, ensuring both original latent spaces are fully spanned by the (private) analysis operator.
The TX applies its analysis operator to the latent vector $\mathbf{x}$ to obtain a semantic code vector:
\begin{equation}\label{eq:semantic_pre_equalizer}
    \c = \F\x=\{\inn{\x}{\f_n}\}_{n=1}^{N} \in \mathbb{C}^N.
\end{equation}
This operation serves as a form of \textit{semantic pre-equalization}, transforming the private latent representation into a shared, relational form that is robust to channel and receiver-specific variations. The resulting vector $ \c $ is then transmitted across the channel, as illustrated in Fig. \ref{fig:semcom_system_model}. Now, thanks to the relative representation principle in (\ref{eq:relrep_similarity}), which assumes that semantic information can be preserved through normalized inner products with respect to the shared anchors, we have the approximate relational consistency:
\begin{align}\label{ass:frame_based_RR}
    \langle \mathbf{x}, \mathbf{f}_n \rangle \approx \langle \mathbf{y}, \mathbf{g}_n \rangle \quad \forall\,n,\ \forall\,\mathbf{s} \in \Dcal,
\end{align}
which links the TX-side projections of $\mathbf{x}$ to the RX-side projections of $\mathbf{y}$ via corresponding anchor-derived directions. Therefore, given a transmitted vector $\c$ in \eqref{eq:semantic_pre_equalizer}, the RX can perform a zero-shot reconstruction of the latent representation as:
\begin{align}\label{eq:zero-shot-sem-eq}
    \hat{\mathbf{y}} \approx \sum_{n=1}^N \underbrace{\langle \mathbf{x}, \mathbf{f}_n \rangle}_{[\c]_n} \mathbf{g}_n, \quad \forall\,\mathbf{s} \in \Dcal,
\end{align}
where each coefficient $[\c]_n$ represents the projection of the TX latent vector onto the $n$-th latent anchor directions. This operation at the RX, which reconstructs a compatible latent vector in its own representation space using only the received semantic code and its local synthesis operator $\{\mathbf{g}_n\}_{\Acal_N}$, is referred to as \textit{semantic post-equalization}, as illustrated in Fig. \ref{fig:semcom_system_model}. Semantic post-equalization enables the RX to interpret transmitted content in a locally meaningful way, achieving seamless semantic alignment without requiring access to the TX encoder, joint training, or additional optimization, thus allowing zero-shot communication between heterogeneous AI-native agents.

\textbf{Remark 3.} The proposed Parseval frame equalizer can be viewed as a special case of the RR framework described in Section~\ref{sec:background:relrep}, where the similarity function is instantiated as the inner product (or equivalently, cosine similarity when the representation space $\mathcal{X}_{(\cdot)}$ is $\ell_2$-normalized). However, it also represents a principled application of the RR framework, particularly in scenarios where the anchor matrix $\mathbf{A}$ is \emph{redundant} ($N > d$). In such cases, reconstructing the original latent space from the relative (anchor-based) representation requires the application of a pseudoinverse (cf. Sec. \ref{sec:background:relrep}). The numerical stability of this operation is governed by the condition number of the Gram matrix $\mathbf{A}^\top\mathbf{A}$. When this matrix is ill-conditioned, even small perturbations—such as quantization noise, channel distortions, or anchor mismatches—can be significantly \emph{amplified}, compromising the robustness of semantic reconstruction. By contrast, enforcing a Parseval frame structure ensures that the frame operator is the identity, yielding a perfectly conditioned pseudoinverse and thus mitigating such instabilities by design. 
\begin{figure*}[!t]
    \centering
    \includegraphics[width=\textwidth, trim=0bp 0bp 0bp 0bp, clip]{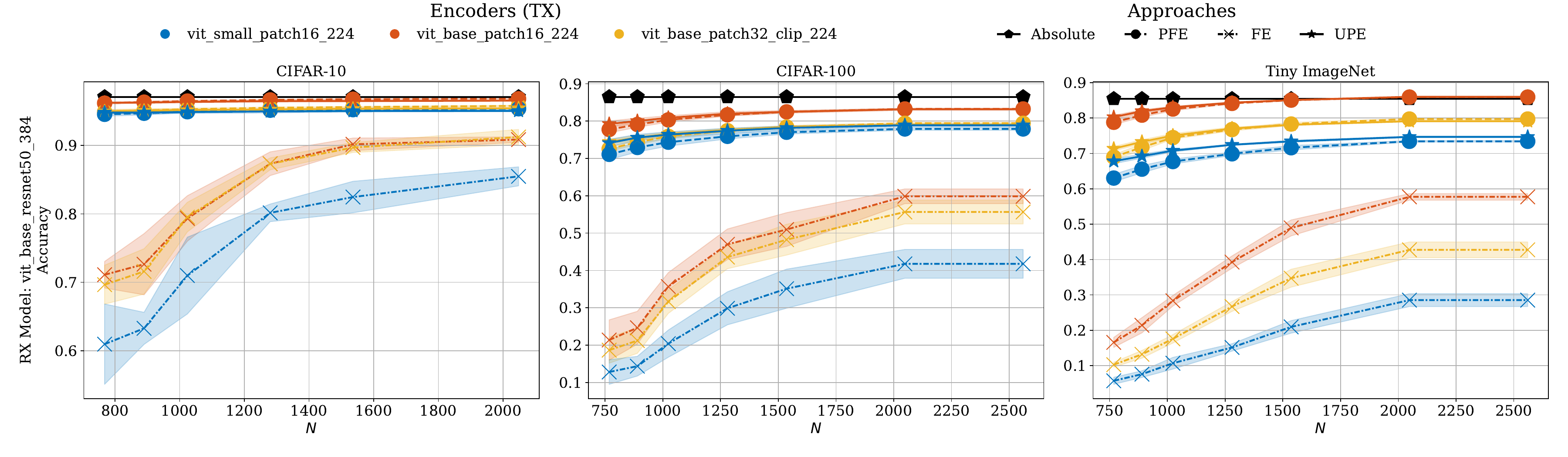}
    \vspace{-0.2cm}
    \caption{
        Task accuracy comparison on CIFAR-10, CIFAR-100, and Tiny-ImageNet using FE, PFE, and the supervised UPE. The Absolute baseline reflects performance with no semantic mismatch. 
    }
    \label{fig:acc_vs_frame}
    \vspace{-0.2cm}
\end{figure*}
 
\textbf{Remark 4.} The inherent alignment capability of RRs is unknown \textit{a priori}, nevertheless, in \eqref{ass:frame_based_RR} we are assuming the existence of a mapping $\Tbf: \X_\Kcal \to \X_\Hcal$ that aligns the two spaces, i.e.,
\begin{align}\label{eq:unit_equality}
    \langle \x, \f_n \rangle \approx \langle \Tbf \y, \Tbf \g_n \rangle = \langle \y, \g_n \rangle.
\end{align}
This condition is satisfied by angle-preserving linear transformations, i.e., $\mathbf{T}\x=\lambda \mathbf{Q}\x$ with $\lambda\neq 0$ and $\mathbf{Q}$ a unitary matrix. With $\lambda=1$ or for normalized spaces, $\mathbf{T}$ reduces to a unitary (or orthogonal) matrix. Although this assumption restricts the framework to cases in which a single global unitary transformation suffices, such cases are nevertheless common in practice, e.g., \cite{gower2004procrustes, zhang2016ten, artetxe2017learning, alvarez2019towards, maiorca2024latent}. This type of alignment, when performed in a supervised setting with access to paired latent representations, corresponds to the well-known (unitary) Procrustes problem \cite{horn2012matrix}:
\begin{align}\label{pr:procustes}
    \min_{\mathbf{P} \in \mathcal{U}_d} \| \mathbf{H} - \mathbf{K} \mathbf{P} \|_F^2,
\end{align}
where $ \mathbf{K}, \mathbf{H} \in \mathbb{R}^{N \times d} $ represent two sets of corresponding latent representations—often referred to as \textit{semantic pilots}—from the transmitter and receiver, respectively. The objective is to find the unitary matrix $ \mathbf{P} $ that best aligns the representations in the least-squares sense. This problem admits a closed-form solution via singular value decomposition (SVD) \cite{schonemann1966generalized}. Specifically, given the SVD of $ \mathbf{K}^\top \mathbf{H} = \mathbf{U} \Sigma \mathbf{V}^\top $, the optimal unitary alignment matrix is
    $\tilde{\mathbf{P}} = \mathbf{U} \mathbf{V}^\top$.
If the latent dimensions of $ \mathbf{K} $ and $ \mathbf{H} $ differ—e.g., $ \mathbf{K} \in \mathbb{R}^{N \times d} $, $ \mathbf{H} \in \mathbb{R}^{N \times p} $ with $ d \neq p $—then the optimal solution $ \tilde{\mathbf{P}} \in \mathbb{R}^{d \times p} $ is semi-orthogonal, preserving orthogonality only along the lower-dimensional subspace.  

\textbf{Results.} We present numerical results to evaluate the performance of the proposed Parseval Frame Equalizer, in comparison with several baseline methods: the Frame Equalizer (corresponding to a plain RR approach \cite{maiorca2024latent}), the supervised Unitary Procrustes Equalizer (UPE) defined in \eqref{pr:procustes}, and the Absolute baseline, which represents an ideal scenario with no semantic mismatch between TX and RX. Experiments consider classification tasks and are conducted on CIFAR-10, CIFAR-100, and Tiny-ImageNet, comprising 10, 100, and 200 classes, respectively. A single MLP-based decoder is trained to interpret the “absolute” representations from the \textit{vit\_base\_resnet\_50\_384} model, pre-trained on ImageNet, for each classification task. Concurrently, we transmit the features encoded by other pre-trained ViT models to the same decoder (see Tab. \ref{tab:model_comparison}). Figure \ref{fig:acc_vs_frame} shows the performance in terms of classification accuracy as a function of the number $N$ of coefficients used in the representation defined in \eqref{eq:semantic_pre_equalizer}, across different datasets, semantic equalization strategies, and encoders. As we can see from Fig. \ref{fig:acc_vs_frame}, PFE consistently achieves higher accuracy than FE, underscoring the advantage conferred by its robust design. Second, increasing the sequence length $N$ of the projected feature vectors yields a modest accuracy gain, likely because a larger and redundant basis attenuates semantic noise during communication. Finally, PFE’s performance converges on that of the supervised UPE. In the UPE setup, we perform alignment using exactly $N$ paired latent vectors as semantic pilots. By holding $N$ constant across FE, PFE, and UPE, we ensure that the number of semantic pilots and the operator size are identical, allowing a fair, sample-wise comparison. Importantly, despite using no semantic pilots, our zero-shot PFE framework nearly matches UPE’s supervised performance on all three datasets.  
\begin{table}[t]
\centering
\begin{tabular}{|lrrrr|}
\hline
\textbf{Model}  & $\!$$\!$$\!$\textbf{\#Param.} & \textbf{FLOPs} & $\!$$\!$$\!$$\!$\textbf{Lat. dim.} &$\!$$\!$$\!$ \textbf{Device} \\ \hline
vit\_small\_patch16\_224 & 21.6M & 8.5 G & 384 & UE-3\\ \hline
vit\_base\_patch16\_224 & 85.7M & 33.72 G & 768 & UE-2 \\ \hline
vit\_base\_resnet50\_384  & 98.1M & 99.04 G & 768 & MEH \\ \hline
vit\_base\_patch32\_clip\_224 & 88.2M & 8.8 G & 768 & UE-1 \\ \hline
\end{tabular}
\caption{Timm models used in the numerical results.} 
\vspace{-0.2cm}
\label{tab:model_comparison} 
\end{table}

\begin{figure*}[!t]
    \centering
    \includegraphics[width=\textwidth]{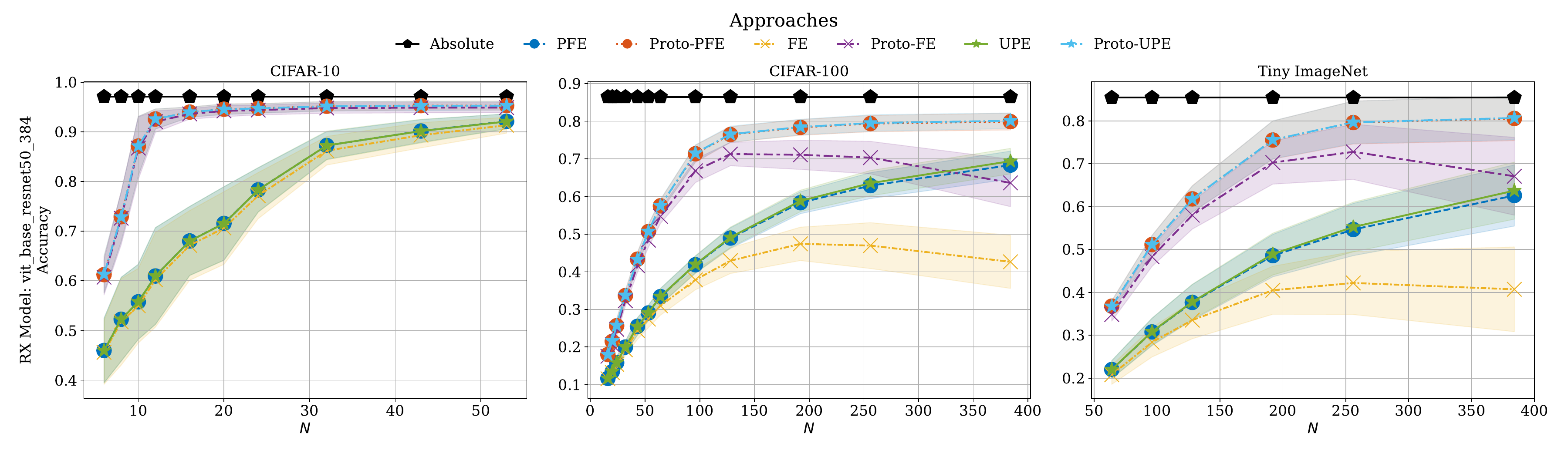}
    \vspace{-0.2cm}
    \caption{
        Average task accuracy over the ensemble of UE encoders under compression for PFE, FE, and the supervised UPE, with and without PAs. 
    }
    \label{fig:acc_vs_compression}
    \vspace{-0.2cm}
\end{figure*}

\subsection{Joint semantic alignment and compression}\label{sub_sec:signal_compression}

In goal-oriented semantic communications, the focus is typically on compression, aiming to select only the most representative features necessary for the downstream task. In this paragraph, we illustrate how the proposed PFE is capable of performing semantic alignment and compression jointly, without a significant loss in terms of goal-oriented information. Specifically, we want to compute a compact semantic code $\c$ in (\ref{eq:semantic_pre_equalizer}), by designing a set of analysis vectors $\{\f_n\}_{n=1}^{N}$, with $N\ll d$, which form a well-suited basis for the approximation of the latent representation $\x$. From the perspective of frame theory, removing a vector from a frame yields either another frame or an incomplete set. In the latter case, the set no longer spans the full space, i.e., $\spann{\F} \subset \H$, and we must consider the orthogonal projection onto $\spann{\F}$, as shown by the following arguments. 

Let us assume working with an incomplete set $\{\f_n\}_{n=1}^{N}$ with the associated matrix $\F$, where $\F^H\F= \S$ is positive-semidefinite and rank-deficient.
Consequently, the inverse does not exist, and we will instead consider the Moore-Penrose pseudoinverse $\S^{\dagger\frac{1}{2}}$, inverting only the nonzero eigenvalues. Thus, we define 
    $\Ft = \S^{\dagger\frac{1}{2}}\F$.
In this case, it holds that $\Ft\Ft^H=\I_N$ due to the construction of $\S^{\dagger\frac{1}{2}}$. However, the relation $\Ft^H\Ft=\I_d$ does not generally hold. To see this, let us compute $\Ft^H\Ft$:
\begin{align}
    \tilde{\S}&=\Ft^H\Ft=\F^H(\S^{\dagger\frac{1}{2}})^H\S^{\dagger\frac{1}{2}}\F=\nonumber\\
    &=\F^H(\S^{\dagger})\F = \F^H(\F^H\F)^{\dagger}\F.
\end{align}
By definition, $\F^H(\F^H\F)^{\dagger}\F$ represents the orthogonal projection onto the row space of $\F$. This projection matrix is equal to $\I_d$ if and only if $\F$ has full column rank. Since $\F$ is incomplete and does not have full column rank, we have $\Ft^H\Ft\neq\I_d$.
In conclusion, it is straightforward to verify that, given these properties, $\St$ is symmetric ($\St=\St^H$) and idempotent ($\St=\St^2$). This implies that $\St$ is an orthogonal projection operator onto $\spann{\F}$, and consequentially, $\Ft$ is a partial isometry (or a semi-orthogonal matrix). This result implies that $\St$ is ill-conditioned, given the row rank deficiency and, instead, $\Ft$ is well-conditioned on the subspace corresponding to its nonzero singular values (with a condition number of 
1). In the sequel, we will refer directly to $\F$ as $\Ft$ and $\S$ as $\St$. 

The results described above lead to the following signal approximation:
\begin{align}\label{eq:compressed_recon}
    \hat{\x}=\P_{\spann{\F}}\x = \sum_{n=1}^N \underbrace{\langle \mathbf{x}, \mathbf{f}_n \rangle}_{[\c]_n} \f_n,
\end{align}
where $\hat{\x}$ is the projection of $\x$ onto the subspace spanned by the set $\{\f_n\}_{n=1}^N$, with $N \!\ll\! d$. Since we are using only a subspace of $\H$ to describe $\x$, \eqref{eq:compressed_recon} highlights the important role of seeking the anchor set $\Acal_N=\{\s_1,...,\s_N\}\subset \Dcal$ whose semantic embeddings $\{\f_n\}_{n=1}^{N}$ (obtained as in Sec.\ref{sub:PFE}) provide the best $N$-dimensional approximation basis.
This requirement is expressed by the minimum-mean-squared-error problem
\begin{align}\label{pr:best_vectors}
    \min_{\Acal_N \subset \Dcal} \mathbb{E}_\x\left\| \x - \sum_{n=1}^N \langle \x, {\f}_n \rangle {\f}_n \right\|^2,
\end{align}
where the expectation is taken over latent vectors $\x$ drawn from the data distribution. The solution to \eqref{pr:best_vectors} identifies the subset of data signals whose latent directions minimize the average distortion between $\x$ and its orthogonal projection onto $\spann{\f_1,\dots,\f_N}$, thereby jointly realizing semantic alignment and compression with minimal task-relevant information loss.
Problem \eqref{pr:best_vectors} is NP-Hard, and it has already been investigated previously. In \cite{moschella2022relative}, the authors conducted a preliminary study on the performance of several anchor selection strategies, considering uniform sampling, k-means, the farthest point sampling (FPS), or the k most frequent words for NLP applications \cite{moschella2022relative, maiorca2024latent}. Instead, in \cite{huttebraucker2024relative}, we proposed an alternative anchor selection strategy based on \textit{Prototypical Anchors} (PAs). This method constructs the anchor matrix using cluster-wise prototypical vectors, defined as the mean of $k$ randomly sampled latent vectors from each cluster, see Alg. \ref{alg:proto_anchors}. 
This approach has several practical advantages: (i) it concentrates the anchor vectors in denser regions of the latent space, (ii) it improves robustness against outliers, (iii) it decouples the anchor set size from the size of the \textit{support set}, $\Scal$, and (iv) it asymptotically converges to the true cluster centroids as $k$ increases. As such, PAs offer an efficient and scalable alternative for constructing informative anchor sets in high-dimensional spaces.
The design of novel sampling strategies is beyond the scope of this work. However, given our focus on task-oriented settings (e.g., classification), where semantic information is often concentrated in high-density latent regions, it is natural to adopt a method that aligns with this structure. For these reasons, we build upon the prototypical anchors introduced in our preliminary work \cite{huttebraucker2024relative}, recognizing them as a suitable and effective method for designing the vectors of the analysis operator.

\begin{figure*}[!t]
    \centering
    \includegraphics[width=\textwidth, trim=0bp 20bp 0bp 0bp, clip]{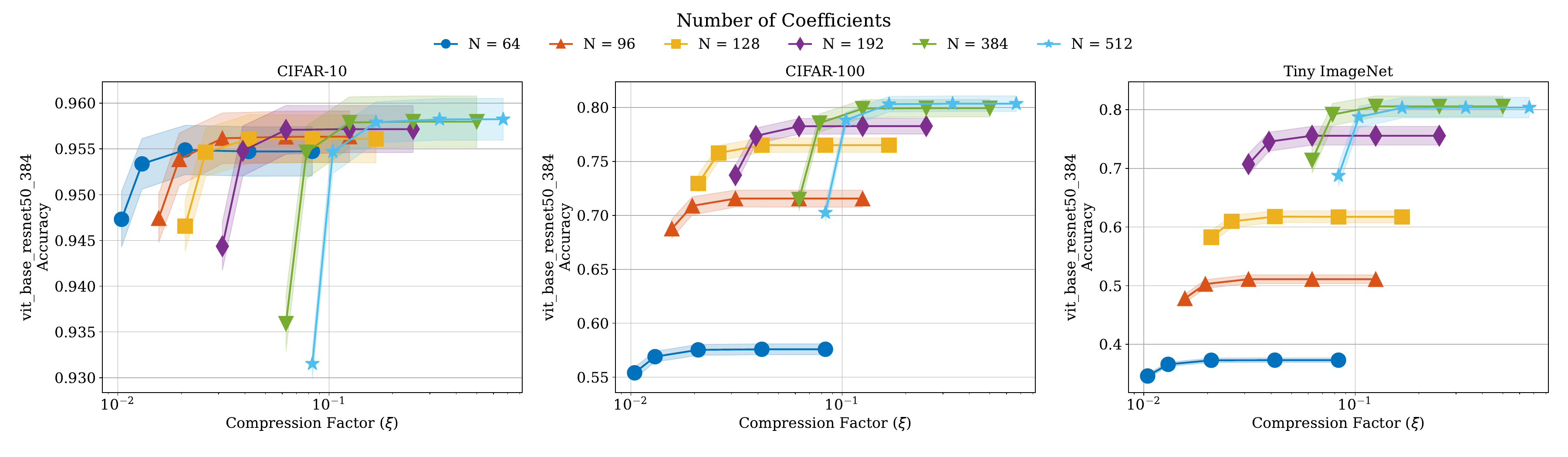}
    \vspace{-0.2cm}
    \caption{
        Task accuracy versus compression factor $\xi$ using Proto-PFE at fixed numbers of transmitted coefficients. Results are averaged over the ensemble of UE encoders. 
    }
    \label{fig:acc_vs_q}
    \vspace{-0.2cm}
\end{figure*}

\begin{algorithm}[t]
\caption{Anchor Prototypes} 
\label{alg:proto_anchors} 
\begin{algorithmic}[1]
\REQUIRE $\Dcal,N,M$ or $\Scal$, $E_\Hcal$
\ENSURE  Anchor matrix $\A$
\IF{$\Scal$ is not given}
    \STATE $\X_\Hcal \gets E_\Hcal(\Dcal)$
    \STATE $\Ccal = \{\Ccal_1, \ldots, \Ccal_{N}\} \gets$ Apply a clustering technique with $N$ clusters to $\X_\Hcal$, $\bigcup_{i=1}^{N} \Ccal_i = \X_\Hcal$
    \STATE $\Scal=\{\Scal_1,\ldots,\Scal_N\} \gets$ For each cluster $\Ccal_i$, draw $M$ samples to form $\Scal_i$
\ENDIF
\STATE Compute the anchor matrix: $\A = \{\a_1, \ldots , \a_N\} \gets \a_i = \frac{1}{M}\sum_{\s\in \Scal_i} E_\Hcal(\s)$
\STATE \textbf{return} $\Scal, \A$
\end{algorithmic}
\end{algorithm}

\textbf{Results.} Figure \ref{fig:acc_vs_compression} mirrors the evaluation in Figure \ref{fig:acc_vs_frame}, now in a compression setting with $N\!<\!d$. We compare the same approaches with and without PAs, and for clarity, report average accuracy over the encoder ensemble.
For the supervised baseline, we also incorporate PAs by deriving a partial isometry $\tilde{\P} = \U\V^H$ through the solution of the Procrustes problem in \eqref{pr:procustes}, aligning the two PA-based analysis operators at the TX and RX. Compression is achieved by truncating the matrices $\U$ and $\V$, whose columns correspond to the principal directions of the aligned subspaces. Specifically, the pre-equalization and post-equalization matrices are defined as $\F_N\! =\! \{\ubf_k\}_{k=1}^N$ and $\G_N \!=\! \{\vbf_k^H\}_{k=1}^N$, where $\ubf_k$ and $\vbf_k$ denote the $k$-th columns of $\U$ and $\V$, respectively.
As shown in Fig. \ref{fig:acc_vs_compression}, PA-based variants consistently outperform their non-PA counterparts across all datasets, underscoring the critical role of support set selection in compressed transmission. On simpler datasets such as CIFAR10, however, the performance gap narrows, suggesting such benchmarks may be insufficiently challenging to fully reveal the advantages of the proposed strategies. In more complex settings, PFE and UPE exhibit nearly identical accuracy, indicating they perform nearly the same transformations. In this context, UPE can be seen as a supervised upper bound that PFE closely approximates in a zero-shot manner, even in the case of joint semantic alignment and compression.

\subsection{The effect of quantization}\label{subsec:quantization}

The semantic code $\c$ must undergo quantization to enable transmission over rate-limited wireless channels. In this paragraph, we investigate the effects of quantizing the transmitted frame coefficients on the semantic alignment and compression capabilities of the proposed method. Since our setting uses normalized vectors and an inner product, each coefficient $[\c]_k$ 
lies in the interval $[-1,1]$, enabling us to employ a fixed-step uniform quantizer. Specifically, we define the step size
  $\Delta \;=\; \frac{2}{2^q - 1},$
where $q$ denotes the number of bits used to represent each coefficient, hence, each 
coefficient is mapped to one of the $2^q$ levels evenly distributed across $[-1,1]$. This rounding process introduces an additive noise term, often 
referred to as \emph{quantization noise}, which degrades the signal quality but can 
yield a significant reduction in payload size. The compression is measured in terms of a \emph{compression factor}
  $\xi \;=\; \frac{N \cdot {q}}{N_{\text{abs}} \cdot {32}},$
where $N$ is the number of transmitted coefficients, each quantized to $q$ bits, and $N_{\text{abs}} \cdot {32}$ represents the size (in bits) of a reference latent vector in $\X_{(\cdot)}$, typically stored as 32-bit floating-point values. 

\textbf{Results.} Figure \ref{fig:acc_vs_q} reports the task-average accuracy across an ensemble of user encoders for different values of the compression factor $\xi$, corresponding to varying quantization levels at a fixed number of transmitted coefficients. The quantization levels considered are ${4, 5, 8, 16, 32}$. We adopt the Proto-PFE method and compute the average accuracy over the ensemble of UE encoders. By varying the compression factor, we observe a clear trade-off between semantic distortion and quantization effects, which impacts goal-oriented performance. Notably, quantization—beyond the number of transmitted representation coefficients—emerges as an additional degree of freedom that can be leveraged to explore the accuracy–compression trade-off. This aspect will be further investigated in the following section, where we address the joint optimization of communication and computation resources for energy-efficient semantic communication under latency and accuracy constraints.

\begin{figure*}[t!]
    \centering
    \includegraphics[width=\textwidth]{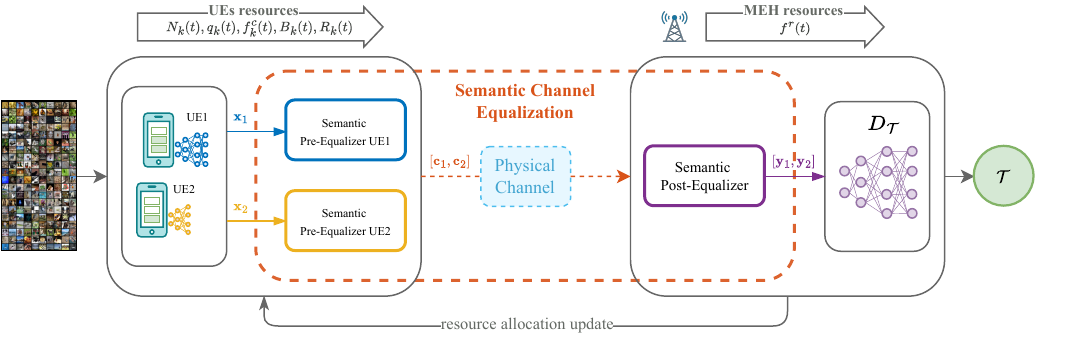}
    \vspace{-.2cm}
    \caption{High-level representation of the multi-agent semantic communication system at time $t$, showing two UEs with private encoders transmitting semantic representations to a MEH.} 
    \label{fig:multi_agent_model_system}
    \vspace{-.2cm}
\end{figure*}

\section{Dynamic Resource Optimization for Frame-based Semantic Channel Equalization and Communication}

In the previous section, we demonstrated how zero-shot SCEq can be achieved using the proposed frame-based approach, highlighting trade-offs between semantic alignment, task performance, and compression. We now build on these results to show how our method can serve as an efficient pre- and post-semantic equalizer and compression scheme, effectively mitigating semantic mismatches arising from diverse DNN-based encoding functions in dynamic multi-agent semantic communication scenarios. Specifically, this section introduces a dynamic optimization strategy designed to adapt system resources---such as communication, computation, and learning---in response to random variations in system parameters, including wireless channel conditions, server availability, and more. We focus on a dynamic resource allocation strategy that minimizes time-average power consumption while satisfying constraints on time-average latency and task accuracy. By dynamically allocating computational resources (e.g., CPU clock frequency), communication resources (e.g., data rate, bandwidth, transmitted coefficients, and quantization level), and learning resources (e.g., anchor set), our approach enables efficient goal-oriented communication and promotes seamless semantic understanding between TXs and RXs.

\subsection{System Model}

We consider an edge-inference scenario in which $K$ AI-native user equipments (UEs) aim to perform semantic communication with an edge server, also referred to as a mobile edge host (MEH). A pictorial example is shown in Fig. \ref{fig:multi_agent_model_system}. The communication process is organized into discrete time slots, indexed by $t$, during which resources are synchronously allocated based on the current system conditions. Each UE $k$ is equipped with a distinct (private) DNN-based encoder, $E_k$, and a family of support sets with different cardinalities, $\{\mathcal{S}_N\}_{N \in \mathcal{N}}$. For convenience, we denote by $N_k(t)$ the number of coefficients transmitted by the $k$-th UE at time $t$, corresponding to the support set $\mathcal{S}_N$. We consider a capacity-achieving digital communication system, assuming perfect Channel State Information (CSI) available at the transmitter. The transmitted message is received by a MEH that employs a decoder $D_{\text{rx}}$, operating within its own latent space $\X_{\text{rx}}$. This reflects a heterogeneous environment, where independently trained encoder schemes optimized on identical tasks and data may lead to semantic mismatches that necessitate SCEq. At each time slot, new inputs are randomly sampled at the UEs, and channel conditions are subject to change. The available total bandwidth $B$ is dynamically partitioned among the UEs such that $\sum_{k=1}^K B_k = B$, where $B_k$ is the bandwidth allocated to user $k$. In the following, we provide a detailed description of the main system parameters.

\paragraph{Power Consumption} 
We consider two primary sources of power consumption for the transmitters, i.e., local computation and local transmission. At time $t$, the power consumed by UE $k$ for local computation is
    $p_k^c(t) = \kappa_k^c (f_k^c(t))^3,$
where $\kappa_k^c$ is the effective switched capacitance of the processors \cite{burd1996processor, brochard2019energy}, and $f_k^c(t)$ is the CPU clock frequency (in Hz). From Shannon capacity formula, given the achievable rate $R_k(t)$ of user $k$, the TX power at time $t$ is given by:
\begin{align}
    p_k^{u}(t) = \frac{B_k(t) N_0}{|h_k(t)|^2} \Bigg[\exp\Bigg(\frac{R_k(t)}{B_k(t)} \ln 2\Bigg) - 1\Bigg],
\end{align}
where $B_k(t)$ is the bandwidth (Hz), $|h_k(t)|^2$ is the uplink channel power gain, and $N_0$ is the power spectral density for thermal noise. At the MEH, the  power consumption required to process the received messages is
    $p^r(t) = \kappa^r (f^r(t))^3,$
where $\kappa^r$ is the effective switched capacitance. Finally, the total power consumption at each time slot is:
\begin{align}\label{eq:total_power}
    p(t) = p^r(t) + \sum_{k=1}^{K} \big(p_k^{u}(t) + p_k^{c}(t)\big).
\end{align}

\paragraph{Latency} The overall latency at the TXs during each time slot has two primary sources, i.e., local processing time and uplink communication time. These are influenced by the encoder design, the number of coefficients used for transmission, the adopted quantization scheme, and communication/computation resources. Specifically, the \textit{local processing time} at the TX is:
\begin{align}
    L^c_k(t) = \frac{C_k^{\text{TX}}(E_k,N_k(t))}{f_k^c(t)},
\end{align}
where $C_k^{\text{TX}}(E_k,N_k(t))$ represents the number of CPU cycles required to compute the absolute representation defined by the encoder $E_k$ and to generate the $N_k(t)$ coefficients for transmission. The \textit{uplink communication time} is calculated as:
\begin{align}
    L^u_k(t) = \frac{N_k(t) \cdot q}{R_k(t)},
\end{align}
where $N_k(t) \cdot q$ is the number of bits of the semantic representation in (\ref{eq:semantic_pre_equalizer}) to be transmitted. Thus, the total latency for the $k$-th agent during a given time slot is:
\begin{align}\label{eq:tx_latency}
    L_k^{TX}(t) = L_k^u(t) + L_k^c(t).
\end{align}
At the receiver, assuming sequential processing of the UE's requests, the \textit{remote processing time} is given by:
\begin{align}
    L^r(t) = \frac{C^{\text{RX}}(t)}{f^r(t)} = \frac{1}{f^r(t)} \sum_{k=1}^K C_k^{\text{RX}}(t),
\end{align}
where $C_k^{\text{RX}}(t)$ denotes the computational load associated with the $k$-th UE request at time $t$, and is defined as
    $C_k^{\text{RX}}(t) = C_{\text{recon}}(N_k(t)) + C_{\text{pred}}(D),$
where $C_{\text{recon}}(\cdot)$ and $C_{\text{pred}}(\cdot)$ represent the number of CPU cycles required for reconstructing the absolute representation and performing the downstream prediction task, respectively.
Finally, the total latency is determined by the maximum communication delay across all agents at time $t$, combined with the remote processing time at the MEH:
\begin{align}\label{eq:total_latency}
L(t) = \max_k \left\{L_k^{TX}(t)\right\} + L^r(t).
\end{align}

\paragraph{Learning Accuracy} We assume that the RX is provided with a validation set $\X^{\textrm{VAL}}$, which is used to evaluate task-dependent accuracy function $G$ to assess learning performance based on encoders $E_k$, number of transmitted coefficients $N_k$, and quantization level $q_k$. As an example, consider a classification task whose validation accuracy can be used to estimate learning performance as
\begin{align}\label{eq:accuracy}
    G(E_k,{N_k(t)},q_k(t)) \!=\! \frac{1}{|\X^{\textrm{VAL}}|} 
    \sum_{x,y  \in \X^{\textrm{VAL}}}\mathbb{I}(\widehat{y}_{E_k,{N_k(t)},q_k(t)}(\mathbf{x}) \!=\! y),
\end{align}
where $\widehat{y}_{E_k,{N_k(t)},q_k(t)}(\mathbf{x})$ is the prediction for the sample $\mathbf{x}$ if $E_k$, $N_k(t)$ and $q_k(t)$ are used as encoder, support set and quantization level, respectively. For simplicity, we will refer to \eqref{eq:accuracy} as $G_k(t)$. Examples of how the performance metric in \eqref{eq:accuracy} is affected by different encoders and varying the transmitted coefficients are illustrated in Fig. \ref{fig:acc_vs_frame} and Fig. \ref{fig:acc_vs_compression}, varying also the quantization level in Fig. \ref{fig:acc_vs_q}.

\subsection{Problem Formulation}

We now formulate the dynamic resource allocation problem for goal-oriented semantic communication. The objective is to learn the optimal policy that, at each time $t$ and agent $k$, allocates the uplink data rates $\{R_k(t)\}_K$, the TX’s CPU cycles $\{f_k(t)\}_K$, the RX’s CPU cycles $\{f^r(t)\}$, the quantization terms $\{q_k(t)\}_K$, and the number of coefficients $\{N_k(t)\}_K$, with the aim of minimizing the long-term average system power consumption in \eqref{eq:total_power}, subject to constraints on the average inference accuracy in \eqref{eq:accuracy} and the average latency in \eqref{eq:total_latency}. Mathematically, the problem can be cast as:
\begin{align}\label{dynamic_resource_allocation}
    \min_{\Psi(t)} \quad & \!\lim_{t \to \infty} \frac{1}{t}\sum_{\tau=0}^{t-1}\left[  \mathbb{E} \{{p}^r(\tau)\} \!+\! \sum_{k=1}^{K}\mathbb{E} \{{p}_{k}^{u}(\tau)\}\!+\!\sum_{k=1}^{K}\mathbb{E} \{{p}_k^{c}(\tau)\}\right] \nonumber\\
\textrm{s.t.} \quad &
(a) \; \lim_{t \to \infty} \frac{1}{t} \sum_{\tau=0}^{t-1} \mathbb{E} \{ L(\tau) \} \leq \bar{L}; \nonumber \\  
\quad & (b) \; \lim_{t \to \infty} \sum_{\tau=0}^{t-1} \mathbb{E} \{ G_k(\tau) \} \geq \bar{G}_k, \quad \forall k;  \nonumber  \\
\conprd
\end{align}
where 
$\Psi(t)$=$[\{N_k(t)\}_k$$,$$ \{q_k(t)\}$$,$$ \{f_k^c(t)\}_k$$,$$ \{B_k(t)\}_k$$,$$ \{R_k(t)\}_k$$,$ $ \{f^r(t)\}]$
represents the set of control variables dynamically optimized at each time slot. The constraints in \eqref{dynamic_resource_allocation} have the following meanings: (a) ensures that the total mean latency does not exceed a threshold $\bar{L}$; (b) ensures the long-term average inference accuracy for the $k$-th device exceeds a given threshold, $\bar{G}_k$; (c-i) impose instantaneous constraints on the resource variables $R_k(t), f_k^c(t)$, $B_k(t)$, $N_k(t)$, and $q_k(t)$. Specifically, $q_k(t)$ and $N_k(t)$ are restricted to finite discrete sets $\mathcal{Q}$ and $\mathcal{N}$, respectively.

\subsection{Algorithm Solution via Stochastic Optimization}

Given the complexity of the long-term optimization problem in \eqref{dynamic_resource_allocation}, particularly due to the unknown statistics of the underlying random variables, we introduce a dynamic algorithmic framework that recasts \eqref{dynamic_resource_allocation} as a sequence of deterministic problems. This framework addresses the long-term average constraints by formulating them as a stability problem, leveraging the tools of stochastic Lyapunov optimization \cite{2010Neely}. First, we introduce two \textit{virtual queues} to monitor the time-average constraints, with their mean-rate stability serving as an indicator of feasibility. We denote by $Z(t+1)$ the virtual queue related to the global latency constraint (a), defined as:
\begin{align}\label{eq:latency_vq}
    Z(t+1) &= \max\{0,\, Z(t) + \epsilon_z \left( L(t) - \bar{L} \right) \}.
\end{align}
Here, $\epsilon_z$ is a positive step size used to control the convergence speed of the algorithm. Then, we introduce the second virtual queue $Q_n(t+1)$ associated with constraint (b), defined as:
\begin{align}
    Q_k(t+1) &= \max\{0,\, Q_k(t) + \epsilon_q \left( \bar{G}_k - G_k(t) \right) \},\label{eq:accuracy_vq}
\end{align}
with $\epsilon_q > 0$.  These virtual queues are then collected into the vector $\bigPhi(t) = [ Z(t), Q_k(t),...,Q_K(t)]$.
Now, we define the \textit{Lyapunov function} as a scalar measure of the queues' congestion level:
\begin{align}\label{eq:lyapunov_function}
    \mathcal{U}(t) = \mathcal{U}(\bigPhi(t)) = \frac{1}{2} \left[ Z^2(t) + \sum_{n=1}^N Q_n^2(t) \right],
\end{align}
and the \textit{one-slot conditional Lyapunov drift}, i.e., the expected change of \eqref{eq:lyapunov_function} given time slot \textit{t},
\begin{align}\label{eq:lyapunov_drift}
    \Delta_t (\mathcal{U}_t) = \mathbb{E} \{ \mathcal{U}_{t+1} - \mathcal{U}_t | \bigPhi(t) \}.
\end{align}
Minimizing \eqref{eq:lyapunov_drift} leads to the mean rate stability of the queues, ensuring constraints (a) and (b) are satisfied. Since the primary goal is to reduce the total power consumption in \eqref{eq:total_power}, we introduce the following \textit{drift-plus-penalty} function \cite{2010Neely}:
\begin{align}\label{eq:drift-plus-penalty}
    \Delta_p(t) =& \Delta_t (\mathcal{U}_t) + V \cdot \mathbb{E} \left\{ p(t) \,\big|\, \bigPhi(t) \right\} = \nonumber \\=& 
    \mathbb{E} \left\{ \mathcal{U}_{t+1} - \mathcal{U}_t + V \cdot  p(t) \,\big|\, \bigPhi(t) \right\},
\end{align}
where $V$ is a weighting parameter that balances the stabilization of the queues with the minimization of the system’s total power. Proceeding as in \cite{2010Neely}, we minimize an upper bound of the drift-plus-penalty function \eqref{eq:drift-plus-penalty}, given by
\begin{align}\label{eq:upper-bound}
\Delta_p(t) \leq &\; \xi + \mathbb{E} \Big\{ Z(t) \left( {(L^r(t) + \sum_{k=1}^KL_k^{TX}(t))} - \bar{L} \right) + \nonumber\\
& + \sum_{k=1}^K Q_k(t) \left( \bar{G} - G_k(t) \right) +\nonumber\\&
+ V\Big ( p^r(t) + \sum_{k=1}^{K} \big(p_k^{u}(t) + p_k^{c}(t)\big) \Big) \,\Big|\, \Phi(t) \Big\}. 
\end{align}
The derivations leading to \eqref{eq:upper-bound} and the value of $\xi$ can be found in the Appendix. For every time slot $t$, we employ a stochastic optimization approach \cite{2010Neely}, ignoring the expectation term by making decisions based on the observation of instantaneous channel realizations. This leads to the solution of a sequence of deterministic per-slot problems at each time $t$ given by:
\begin{align}\label{prob:min-drift}
\min_{  \bigPsi(t) } \quad& \Gamma(\Psi(t))=\sum_{k=1}^K \Big[ Z(t)L_k^{\text{TX}}(t) - Q_k(t) G_k(t)+\nonumber\\& 
+V( {p}_{k}^{u}(t)+{p}_k^{c}(t)) \Big] + V p^r(t)+ Z(t)L^r(t) \nonumber\\
\textrm{s.t.} \quad \conprd
\end{align}
Due to the dependence of \eqref{prob:min-drift} on $\{N_k(t)\}_k$ and $\{q_k(t)\}_k$, solving the problem directly results in a mixed-integer nonlinear optimization problem. However, by fixing all the $N_k(t)$ and $q_k(t)$, the problem becomes separable into three independent convex sub-problems: 
(i) optimizing the CPU frequencies of UEs, (ii) computing the optimal edge server CPU frequency, and (iii) jointly allocating data rates and bandwidth for UE's communications. Each of these sub-problems allows for closed-form or computationally efficient solutions. 
In the sequel, we detail the formulation and derivation of closed-form solutions for the three sub-problems.

\subsubsection{UEs Computing Resource Allocation}\label{sub:UEsAllo}

UEs computing resource allocation optimizes ${f_k^c(t)}_{k=1}^K$ at each time $t$. From \eqref{prob:min-drift}, we can obtain the following sub-problem:
\begin{align}\label{pr:TX-freq}
    \min_{f^c_k(t)} \quad & Z(t)\frac{C_k^{\text{TX}}(E_k,N_k(t))}{f^c_k(t)} + V \kappa^c_k({f^c_k(t)})^3 \nonumber\\
    \textrm{subject to} \quad
    \constraintend{d}{f_k^{\textrm{MIN}} < f_k^c(t) \leq f_k^{\textrm{MAX}}}{\forall\, k;}
\end{align}
The Lagrangian function of \eqref{pr:TX-freq} writes as:
\begin{align}
    \mathcal{L}_{f^c_k(t)} = Z(t)\frac{C_k^{\text{TX}}(E_k,N_k(t))}{f^c_k(t)} + V \cdot \kappa^c_k({f^c_k(t)})^3\nonumber \\  
    - \delta(f^c_k(t) - f_k^{\textrm{MIN}}) + \nu(f^c_k(t) + f_k^{\textrm{MAX}})
\end{align}
where $\delta$ and $\nu$ are the Lagrange multipliers associated with the constraints of \eqref{pr:TX-freq}. Then, solving the Karush-Khun-Tucker (KKT) conditions \cite{boyd2004convex} of the strictly convex problem \eqref{pr:TX-freq}, we obtain:
\begin{align}\label{eq:optimal_eu_freq}
    {\tilde f^c_k(t)} = \Big[ \Big(\frac{Z(t)\cdot C_k^{\text{TX}}(E_k,N_k(t))}{3\kappa_k^cV}\Big)^{\frac{1}{4}} \Big]^{f_k^{\textrm{MAX}}}_{f_k^{\textrm{MIN}}} \quad \forall\,k,
\end{align}
where the thresholding operator $[x]^b_a := \min\left(\max(x, a), b\right)$ ensures that the computed CPU frequency is projected onto the feasible interval $[f_k^{\textrm{MIN}}, f_k^{\textrm{MAX}}]$.

\subsubsection{MEH Computing Resource Allocation}

The MEH computing resource allocation problem aims to optimize the CPU frequency ${f^r(t)}$. By isolating the corresponding terms in the per-slot objective \eqref{prob:min-drift}, we formulate the following sub-problem:
\begin{align}\label{prob:MEHsub-problem}
    \min_{f^r(t)} \quad & Z(t)\frac{C^{RX}(t)}{{f^r}(t)} + V \kappa^r({f^r(t)})^3 \nonumber\\
    \textrm{subject to} \quad
    \constraintend{d}{{f^r}^{\textrm{MIN}} < {f^r}(t) \leq {f^r}^{\textrm{MAX}}}{\forall\, k.}
\end{align}
Problem \eqref{prob:MEHsub-problem} is a strictly convex optimization problem, which enjoys a simple closed-form solution. Then, solving the KKT conditions, we obtain the following optimal MEH CPU frequency,
\begin{align}\label{eq:optimal_meh_freq}
    {\tilde f^r(t)} = \left[ \Big(\frac{Z(t)\cdot C^{RX}(t)}{3\kappa^rV}\Big)^{\frac{1}{4}} \right]^{{f^r}^{\textrm{MAX}}}_{{f^r}^{\textrm{MIN}}}.
\end{align}

\subsubsection{UE's Data Rate and Bandwidth Allocation}

For each time slot $t$, the data rate and bandwidth allocation sub-problem can be written as follows,
\begin{align}\label{prob:joint_band_rate}
    \min_{R_k(t),B_k(t)} \quad & \sum_{k=1}^{K} Z(t)\frac{N_k(t) \cdot q_k(t)}{R_k(t)} + \nonumber \\& 
    V\cdot \frac{B_k(t) N_0}{|h_k(t)|^2}\Bigg[\exp\Bigg(\frac{R_k(t)}{B_k(t)}\ln2\Bigg)-1\Bigg] \nonumber \\
    \textrm{subject to} \quad
    \constraint{c}{R_k^{\textrm{MIN}} \leq R_k(t) \leq R_k^{\textrm{MAX}}(t)}{\forall\, k,\, t;}\\
    \constraint{g}{B_k(t)\geq B_k^{\text{MIN}}}{\forall\, t,k;}\\
    \constraintend{h}{\sum_{k=1}^{K} {B_k}(t) = {B}}{\forall\, t;}
\end{align}
where the maximum data rate reads as:
\begin{align}
    R_k^{\textrm{MAX}}(t) = B_k(t) \log_2 \left ( 1 + \frac{P^{\text{MAX}}_k|h_k(t)|^2}{B_k(t)N_0} \right ), 
\end{align}
being $P^{\text{MAX}}_k$ the maximum transmission power at time $t$ for the $k$-th UE.
An off-the-shelf convex solver can address \eqref{prob:joint_band_rate}, as the formulation is jointly convex in $\{R_k(t)\}_k$ and $\{B_k(t)\}_k$. However, implementing such a solver for large-scale, real-time applications with many UEs might be prohibitively expensive in terms of computation. Moreover, the presence of multiple system parameters might introduce numerical instabilities that degrade solution accuracy and feasibility in practice. To mitigate these issues, we propose a simple yet effective heuristic that separates the bandwidth allocation from the rate-optimization step, thereby reducing complexity and improving robustness.
Specifically, we allocate each user's bandwidth $ B_k(t) $ proportionally to the product of its number of coefficients $ N_k(t) $ and quantization level $ q_k(t) $, capturing the relative significance of each user's data. The allocation is given by:
\begin{align}\label{eq:sub_optimal_bandwidth}
    \tilde{B}_k(t) = B \cdot \frac{w_k(t)}{\sum_{k=1}^K w_k(t)},
\end{align}
where the weight $ w_k(t) $ is defined as:
\begin{align}\label{eq:weight_factor}
    w_k(t) = \big(N_k(t)\big)^\alpha \cdot \big(q_k(t)\big)^\beta.
\end{align}
Here, the exponents $ \alpha $ and $ \beta $ are tunable parameters that control the relative importance of the number of coefficients $ N_k(t) $ and the quantization level $ q_k(t) $, respectively. This weighted allocation strategy reflects the intuition that users with more coefficients (e.g., larger data volume) or higher quantization levels (e.g., requiring finer precision) should receive a proportionally larger share of the total bandwidth $ B $ to meet their transmission needs more effectively. With $\{\tilde{B}_k(t)\}_k$ given by (\ref{eq:sub_optimal_bandwidth}), problem  \eqref{prob:joint_band_rate} decouples into independent rate-allocation sub-problems, each admitting the closed-form solution:
\begin{align}\label{eq:optimal_data_rate}
    \tilde{R}_k(t)\!=\!\!\left[ \!\frac{2\tilde{B}_k(t)}{\ln 2}W\!\!\left(\! \frac{1}{2\tilde{B}_k(t)}\sqrt{\frac{Z(t)N_k(t)q_k(t)|h_k(t)|^2\ln 2}{VN_0}} \right) \!\right]^{{R}^{\text{MAX}}_k(t)}_{{R}^{\text{MIN}}_k(t)}\!\!\!\!,
\end{align}
$k=1,\ldots,K$, where $W(\cdot)$ is the Lambert function. 

\subsubsection{Dynamic Selection of Transmitted Coefficients and Quantization Level}
\label{sub:Nq-selection}

Having obtained closed-form expressions for the continuous variables
$\{\tilde f_k^{c}(t)\}_{k=1}^K$ \eqref{eq:optimal_eu_freq},
$\tilde f^{r}(t)$ \eqref{eq:optimal_meh_freq},
$\{\tilde B_k(t)\}_{k=1}^K$ \eqref{eq:sub_optimal_bandwidth}, and $\{\tilde R_k(t)\}_{k=1}^K$ \eqref{eq:optimal_data_rate}, the per-slot drift-plus-penalty \eqref{prob:min-drift} reduces to an integer program that depends only on the discrete variables $\{N_k(t)\}_{k=1}^K$ and $\{q_k(t)\}_{k=1}^K$:
\begin{align}
\label{prob:integer-program}
\min_{  \Theta(t)=\{N_k(t),q_k(t)\}_{k=1}^K } \quad & \Gamma(\Theta(t)) 
\end{align}
The naïve exhaustive search scales as $\bigl(|\mathcal N|\times|\mathcal Q|\bigr)^{K}$ and is quickly impractical even for moderate $K$. For this reason, we adopt an iterative greedy procedure that updates one user at a
time while keeping the other $K\!-\!1$ users fixed.  At each iteration, the scheduler evaluates a candidate pair
$(N_k^{\mathrm{cand}},q_k^{\mathrm{cand}})\in\mathcal N\times\mathcal Q$, and evaluates $\Gamma(\Theta(t))$ considering the closed-form solutions described above, i.e., \eqref{eq:optimal_eu_freq},  \eqref{eq:optimal_meh_freq}, \eqref{eq:sub_optimal_bandwidth}, and \eqref{eq:optimal_data_rate}. After exhausting all candidate pairs for user $k$, we select the pair that yields the smallest cost function $\Gamma(\Theta(t))$. We repeat this process for each user $k$. The resulting complexity is $\bigl(|\mathcal N|\times|\mathcal Q|\times K\bigr)$, i.e., linear in the number of users.
Moreover, because the objective function and feasible set in \eqref{prob:integer-program} are bounded for all $t$, the greedy approach is a \emph{C-additive approximation} \cite{2010Neely}: it always remains within a bounded additive gap from the global optimum, making the approach well-suited for dynamic resource allocation.

\medskip
The continuous sub-problems yield closed-form updates for computing and radio resources, while the remaining discrete choices for each UE are refined by the greedy search above. Putting these pieces together, Algorithm \ref{alg:dynamic_resource_allocation} summaries the complete per-slot procedure.

\begin{algorithm}[t]
\caption{Dynamic Resource Allocation Algorithm}
\label{alg:dynamic_resource_allocation}
\begin{algorithmic}[1]
\REQUIRE 
Channel gains $\{h_k(t)\}_{k=1}^K$, bandwidth $B$, sets $\mathcal{N}$, $\mathcal{Q}$; \\
Initial values: $Z(0)$, $\{Q_k(0)\}_{k=1}^K$, $\{N_k(0)\}_{k=1}^K$, $\{q_k(0)\}_{k=1}^K$; \\
Parameters: $\alpha$, $\beta$, $\gamma$, $\epsilon_z$, $\epsilon_q$;
\ENSURE $\{N_k(t)\}_{k=1}^K$,$\{q_k(t)\}_{k=1}^K$,$\{R_k(t)\}_{k=1}^K$,$\{B_k(t)\}_{k=1}^K$, $\{f_k^c(t)\}_{k=1}^K$,$f^r(t)$.
\FOR{each time slot $t \geq 0$}
    \STATE Observe $h_k(t)$ for all UEs; Update $Z(t)$ and $\{Q_k(t)\}_{k=1}^K$ as per \eqref{eq:latency_vq} and \eqref{eq:accuracy_vq};
        \FOR{each user $k = 1$ to $K$}
            \STATE Fix parameters $\{(N_j(t), q_j(t))\}_{j \neq k}$;
            \STATE Set best cost $\Gamma^* \leftarrow +\infty$;
            \FOR{each candidate pair $(N_k^{\text{cand}}, q_k^{\text{cand}}) \in \mathcal{N} \times \mathcal{Q}$}
                \STATE Compute ${f^r}^{\text{cand}}(t)$ via \eqref{eq:optimal_meh_freq} and ${f^c_k}^{\text{cand}}(t)$ via \eqref{eq:optimal_eu_freq} $\forall\,k$;
                \STATE Compute ${B}^{\text{cand}}_k(t)$ via \eqref{eq:sub_optimal_bandwidth} 
                and ${R}^{\text{cand}}_k(t)$ via \eqref{eq:optimal_data_rate} $\forall\,k$;
                \STATE Evaluate cost $\Gamma(\Theta(t))$ as in \eqref{prob:integer-program};
                \IF{$\Gamma(\Theta(t)) < \Gamma^*$}
                    \STATE $(N_k(t), q_k(t)) \leftarrow (N_k^{\text{cand}}(t), q_k^{\text{cand}}(t))$; 
                    \STATE $({f^c_k}(t),{f^r}(t),B_k(t), R_k(t)) \leftarrow ({f^c_k}^{\text{cand}}(t),{f^r}^{\text{cand}}(t),B_k^{\text{cand}}(t), R_k^{\text{cand}}(t))$, $\forall\,k$;
                    \STATE Update best cost: $\Gamma^* \leftarrow \Gamma(\Psi(t))$;
                \ENDIF
            \ENDFOR
        \ENDFOR
    \STATE \textbf{Data Transmission:} Transmit $\{\c_k\}_{k=1}^K$.
\ENDFOR
\end{algorithmic}
\end{algorithm}

\subsection{Numerical Results}

\begin{figure*}[!t]
    \centering
    \begin{minipage}{0.45\textwidth}
        \centering
        \includegraphics[width=\linewidth]{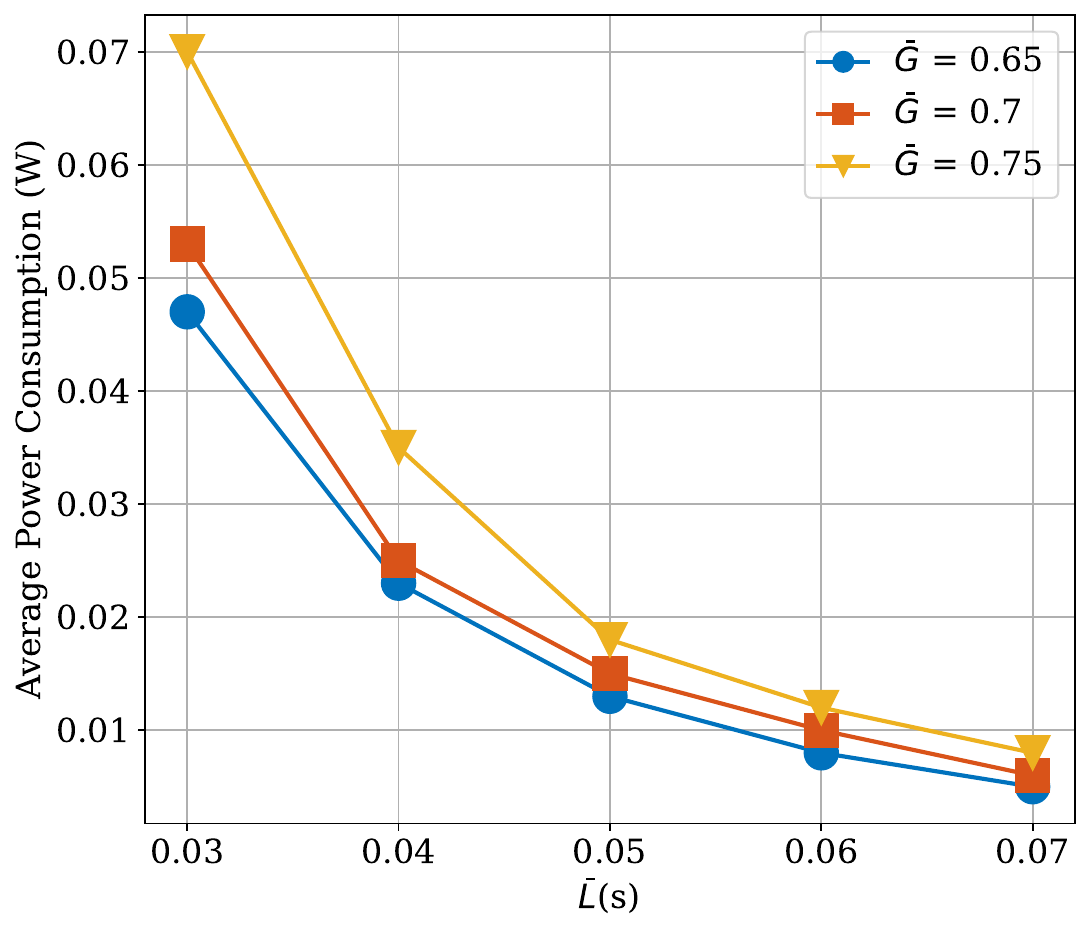}
    \end{minipage}
    \hfill
    \begin{minipage}{0.45\textwidth}
        \centering
        \includegraphics[width=\linewidth]{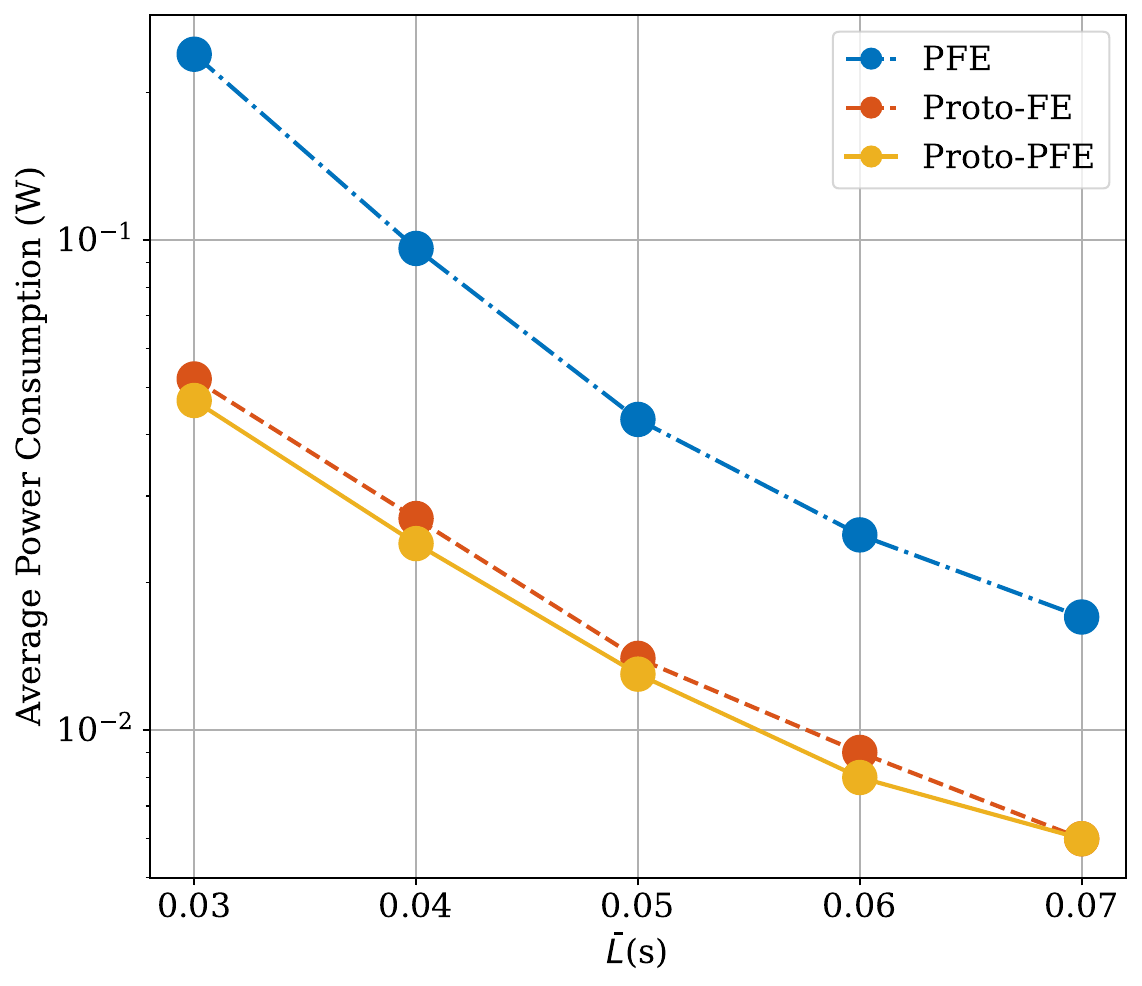}
    \end{minipage}
    \caption{Left: Average power consumption with Proto-PFE vs. latency and accuracy targets. Right: Power consumption with fixed $\bar{G}=0.65$ using Proto-PFE, Proto-FE, and PFE.}
    \label{fig:trade-off-curve}
\end{figure*}

\begin{figure*}[!t]
    \centering
    \begin{minipage}{0.45\textwidth}
        \centering
        \includegraphics[width=\linewidth]{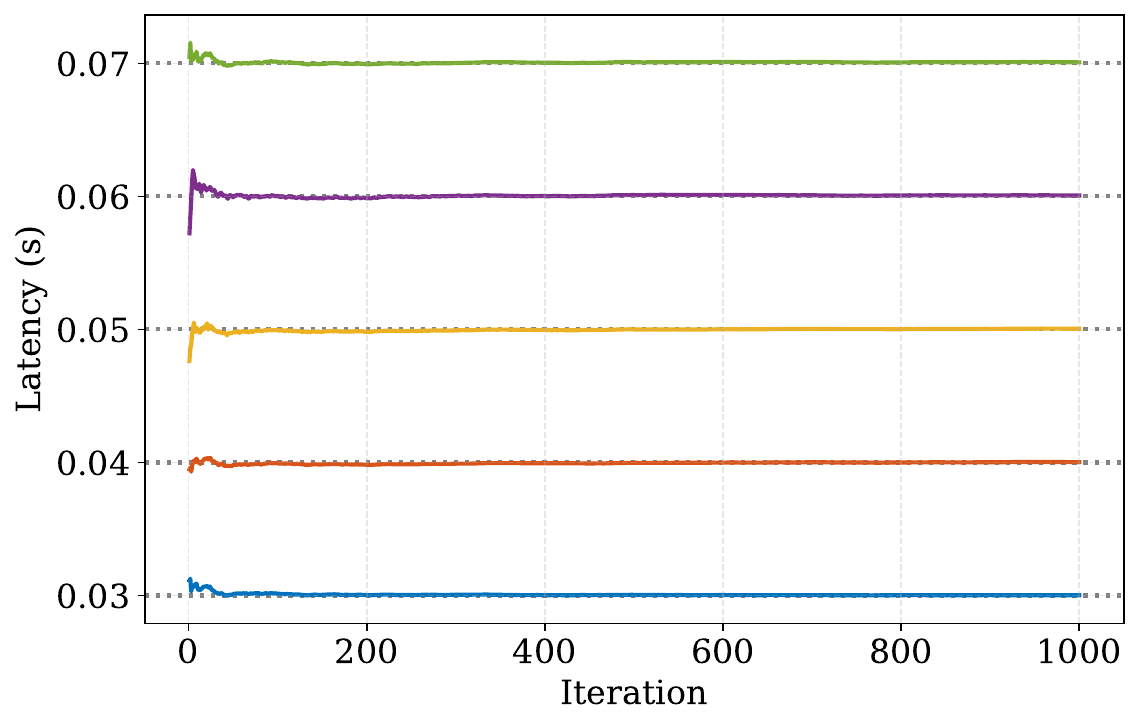}
    \end{minipage}
    \hfill
    \begin{minipage}{0.45\textwidth}
        \centering
        \includegraphics[width=\linewidth]{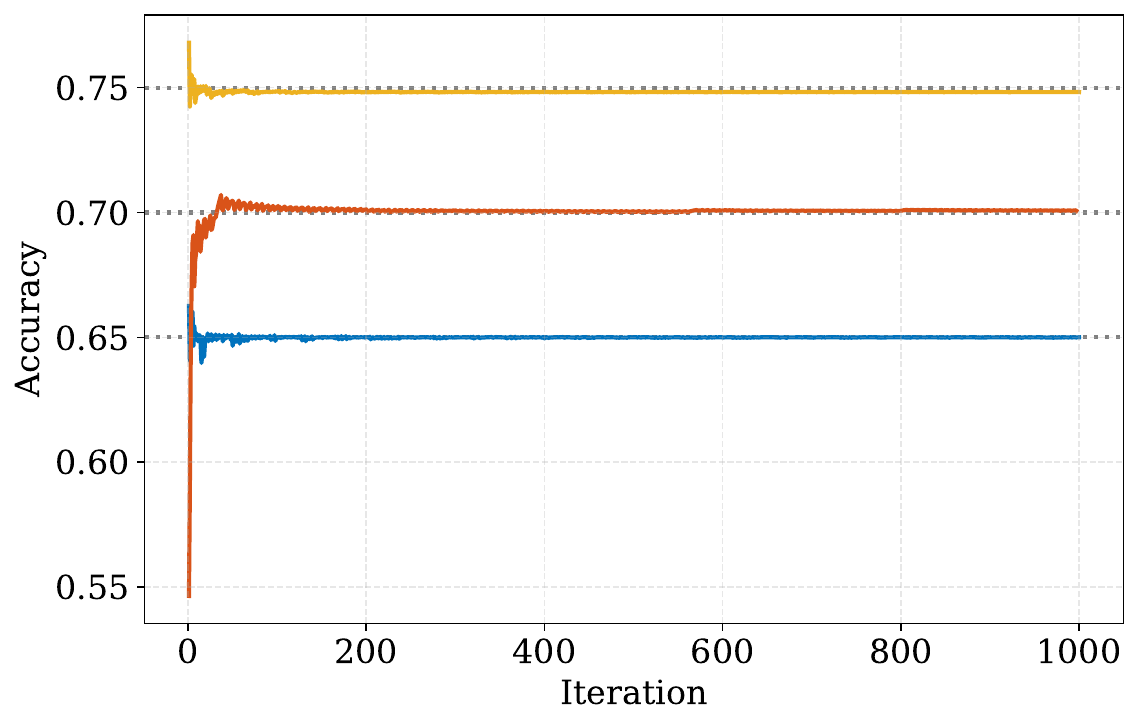}
    \end{minipage}
    \caption{Running averages of latency (left) and accuracy (right) queues for varying targets, with fixed $\bar{G} = 0.7$ and $\bar{L} = 0.04$, respectively.}
    \label{fig:target-queue-curve}
\end{figure*}

To evaluate our method, we execute Algorithm \ref{alg:dynamic_resource_allocation} varying target latency ($\bar L$) and target accuracy ($\bar G$), to characterize the long‑term trade‑off between average power consumption, inference latency, and classification accuracy. The system includes three UEs, each running a different pre-trained DNN encoder via the Timm library \cite{rw2019timm}. The MEH uses the \textit{vit\_base\_resnet50\_384} encoder, while the three transmitters employ \textit{vit\_small\_patch16\_224}, \textit{vit\_base\_patch16\_244}, and \textit{vit\_base\_patch32\_clip\_224} (as stated in Tab. \ref{tab:model_comparison}). At the MEH, an MLP classifier performs 200‑class classification on the Tiny‑ImageNet dataset \cite{tinyimagenetHF}.
We discretise transmitted coefficients and quantization bits as $\Ncal = \{32,64,96,128,192,384,512\}$, and $\Qcal = \{2,4,6,8,12,16,32\}$.
The control parameters $V$, $\epsilon_z$, $\epsilon_q$, and the algorithm hyperparameters in \eqref{eq:weight_factor}, $\alpha$ and $\beta$, are selected via grid search for each $(\bar L,\bar G)$ combination. Each run uses $7.5\times10^3$ iterations.

For this setting, we set the radio and computation parameters as follows: $f_k^{\text{max}}=3.5$GHz, ${f^r}^{\text{max}}=4$GHz, the total bandwidth is $B=500$kHz, 
$P_k^{\text{max}}=0.15$W and $P_{\text{MEH}}^{\text{max}}=0.2$W.  
The noise spectral density is $N_0=k_B T$, with Boltzmann constant $k_B$ and temperature $T=290$K.
We model small‑scale fading as Rayleigh (unit variance) with free‑space path loss, $P_L(dB)= 20\log_{10}(d)+20\log_{10}(f)+92.45$, using distance $d=0.1$km and carrier frequency $f=3.5$GHz. 

Figure \ref{fig:trade-off-curve}-(left) shows the average power consumption over the final $10^3$ iterations for each combination of latency target $\bar L$ and accuracy target $\bar G$. The curves clearly demonstrate the trade‑off enabled by the proposed dynamic resource allocation methodology using Proto-PFE: relaxing the latency constraint reduces average power consumption for all accuracy levels, while lowering the accuracy target further decreases power usage at a given latency. Conversely, achieving higher accuracy requires increased power, especially under tighter latency requirements. These results highlight the flexibility in balancing energy efficiency, inference speed, and task performance.
\begin{figure*}[t!]
    \centering 
    \includegraphics[width=0.9\textwidth]{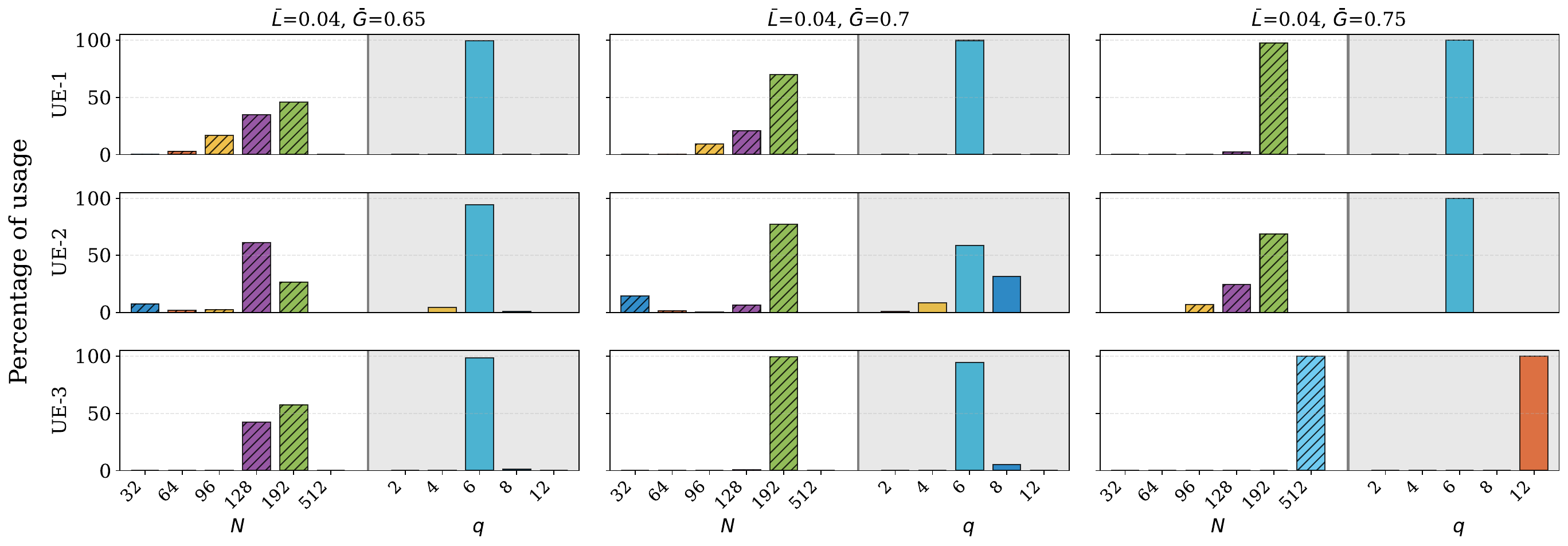}
    \caption{Empirical distribution of selected transmission parameters per UE (rows) under a fixed target latency $\bar{L} = 0.04$ and increasing target accuracy $\bar{G}$ (0.65, 0.70, 0.75). Bars to the left of the vertical line show the usage frequency of coefficient counts $N\! \in\! \{32,64,96,128,192,512\}$, while bars to the right indicate the percentage of use of quantization levels $q\! \in\! \{2,4,6,8,12\}$.}
    \label{fig:Nq_distro}
\end{figure*}

Figure \ref{fig:trade-off-curve}-(right) focuses on the case $\overline{G}=0.65$ and compares PFE, Proto-FE, and Proto-PFE, underlying latency budgets.
This accuracy level is deliberately moderate: for tighter accuracy targets, PFE and Proto-FE are already unable to satisfy the constraint with Tiny-Imagenet (see Fig. \ref{fig:acc_vs_frame} and \ref{fig:acc_vs_compression}).
To give those baselines the widest margin, we extended their coefficient budget to $\mathcal{N}=\{...,512,1024,1536,2048\}$.
Even with this enlarged design space, PFE remains substantially less energy-efficient across the entire latency range. Proto-PFE attains the lowest power consumption and almost overlaps the Proto-FE curve. This small residual gap reflects the relaxed accuracy target and the close performance reached in Fig. \ref{fig:acc_vs_compression} until $128$ coefficients. 

In Figure \ref{fig:target-queue-curve}, we show how the proposed methodology with Proto-PFE satisfies the long-term average latency and accuracy constraints. Specifically, in the left plot, we fix the accuracy target, $\overline{G} = 0.7$, and vary the latency target. In the right plot, we fix the latency target, $\overline{L}$, and vary the accuracy target. The left plot shows the running average of latency, while the right plot shows the running average of accuracy, each superimposed on horizontal dashed lines indicating the corresponding threshold value, demonstrating that the system consistently meets the specified constraints. In Fig. \ref{fig:Nq_distro}, we plot the empirical distribution of the selected transmission parameters $(N,q)$ for each UE with accuracy constraint $\bar G$ equal to  $\{0.65, 0.7, 0.75\}$, with latency bound fixed at $\bar L = 0.04$. As expected, when the target accuracy $\bar{G}$ increases, the algorithm shifts toward configurations with higher aggregate bit-loads ($N \times q$). Conversely, as $\bar{G}$ is relaxed, it reduces the bit-load, demonstrating its ability to adaptively balance the number of coefficients and quantization resolution to meet varying accuracy requirements under a fixed latency budget.

\section{Conclusion}

In this work, we proposed Parseval Frame Equalizer, a semantic channel equalization scheme for goal-oriented semantic communication systems. PFE leverages RRs integrated with a frame-based approach to align latent spaces at the receiver, enabling an efficient semantic equalizer design. It dynamically manages the trade-off between signal compression and expansion, using lightweight dot product operations, achieving low overhead, robust, and zero-shot alignment of latent spaces at the receiver. Comparisons between equalizer methods show that the Parseval-based solution yields enhanced stability and improved performance over competing approaches, while experiments on signal compression and quantization reveal the inherent trade-offs between semantic and quantization noise, highlighting the importance of the selected anchor set.

We further integrated our approach into a dynamic optimization framework based on Lyapunov stochastic optimization, which jointly allocates communication, computation, and learning resources in a multi-agent setting. Extensive numerical experiments demonstrate that our system meets long-term latency and accuracy constraints, along with balances energy efficiency, inference speed, and task performance under diverse operating conditions. Future work may address the overhead associated with transmitting semantic pilot sets and explore more advanced sharing strategies to further enhance alignment robustness under challenging network conditions. Overall, our results pave the way for energy-efficient, low-latency semantic communication systems that can adapt to the complexities of heterogeneous AI-Native wireless networks.

\bibliographystyle{IEEEtran}
\bibliography{ref}

\end{document}


\section{Appendix}

\subsection{Upper Bound of (28)}

We now derive a convenient upper bound on the drift-plus-penalty term:
\begin{align}
    \Delta_p(t)=&\mathbb{E}\Bigl\{\tfrac12\bigl(Z^2(t+1)-Z^2(t)\bigr)+\nonumber\\
    +&\tfrac12\sum_{n=1}^N\bigl(Q_n^2(t+1)-Q_n^2(t)\bigr)
    +V\,p(t)\,\Bigm|\,\Phi(t)\Bigr\}.
\end{align}
\\
\textbf{Bounding the latency terms.}
Let define $\Delta_Z \;=\;\tfrac12\bigl(Z^2(t+1)-Z^2(t)\bigr)$. Using the latency virtual queue in (24)
, we obtain
\begin{align}\label{eq:latency_diff_app}
  \Delta_Z
  &\le \epsilon_z^2\bigl(L(t)-\bar L\bigr)^2
      +2\,\epsilon_z\,Z(t)\bigl(L(t)-\bar L\bigr)\\
  &\le \epsilon_z^2(L^{\max}-\bar L)^2
      +2\,\epsilon_z\,Z(t)\Bigl(\max_k\{L_k^{TX}(t)\}+L^r(t)-\bar L\Bigr)\nonumber\\
  &\le \epsilon_z^2(L^{\max}-\bar L)^2
      +2\,\epsilon_z\,Z(t)\Bigl(\sum_{k=1}^K L_k^{TX}(t)+L^r(t)-\bar L\Bigr),
\end{align}
where we used that $(L(t)-\bar L)^2\le(L^{\max}-\bar L)^2$ and $\max_kL_k^{TX}(t)\le\sum_kL_k^{TX}(t)$.
\\
\textbf{Bounding the accuracy terms.}
Similarly, for each accuracy queue $Q_k(t)$ with update (25)
, let
$\Delta_{Q_k} \;=\;\tfrac12\bigl(Q_k^2(t+1)-Q_k^2(t)\bigr)$.
Since $G_k(t)\le G_k^{\max}$, we get
\begin{align}\label{eq:accuracy_diff_app}
  \Delta_{Q_k}
  &\le \epsilon_q^2\bigl(\bar G_k - G_k(t)\bigr)^2
      +2\,\epsilon_q\,Q_k(t)\bigl(\bar G_k - G_k(t)\bigr)\nonumber\\
  &\le \epsilon_q^2\bigl(\bar G_k - G_k^{\max}\bigr)^2
      +2\,\epsilon_q\,Q_k(t)\bigl(\bar G_k - G_k(t)\bigr).
\end{align}

Summing \eqref{eq:latency_diff_app} over $Z$ and \eqref{eq:accuracy_diff_app} over $k$, then adding the penalty term, yields
\begin{align}
  \Delta_p(t)
  \le \xi &+ \mathbb{E}\Bigl\{
    Z(t)\bigl(L^r(t)+\sum_{k=1}^K L_k^{TX}(t)-\bar L\bigr)+\nonumber\\
  &+\sum_{k=1}^K Q_k(t)\bigl(\bar G_k - G_k(t)\bigr)+\nonumber\\
  &+V\Bigl(p^r(t)+\sum_{k=1}^K\bigl(p_k^u(t)+p_k^c(t)\bigr)\Bigr)
    \,\Bigm|\,\Phi(t)\Bigr\},
\end{align}
where
\begin{align}
  \xi = \tfrac12\,\epsilon_z^2(L^{\max}-\bar L)^2
        +\tfrac12\sum_{k=1}^K\epsilon_q^2\bigl(\bar G_k - G_k^{\max}\bigr)^2.
\end{align}                   